\documentclass[12pt]{article}
\usepackage{stmaryrd}
\usepackage{amssymb}
\usepackage{graphics}
\usepackage{epsfig}
\usepackage{a4wide}

\begin{document}
\newcommand{\qqttz}{$q\bar{q} \to t \bar t Z^0~$ }
\newcommand{\ggttz}{$gg \to t \bar t Z^0~$ }
\newcommand{\qqttzg}{$q\bar{q} \to t \bar t Z^0 g~$ }
\newcommand{\qgttzq}{$q(\bar{q})g \to t \bar t Z^0 q (\bar{q})~$ }
\newcommand{\qqggttz}{$q\bar{q}(gg) \to t \bar t Z^0$ }
\newcommand{\qqggttzg}{$q\bar{q} (gg) \to t \bar t Z^0 g~$ }
\newcommand{\ppqqttz}{$pp \to q\bar q \to t\bar t Z^0 +X~$}
\newcommand{\ppuuttz}{$pp \to u\bar u \to t\bar t Z^0 +X~$}
\newcommand{\ppggttz}{$pp \to gg \to t\bar t Z^0 +X~$}
\newcommand{\ppttz}{$pp \to t\bar t Z^0 +X~$}

\title{ Supersymmetric QCD and CP-violation effects in $t \bar t Z^0$ production
at the LHC    }
\author{ Liu Ning, Guo Lei, Ma Wen-Gan, Zhang Ren-You, and Han Liang   \\
{\small Department of Modern Physics, University of Science and Technology  }  \\
{\small of China (USTC), Hefei, Anhui 230026, P.R.China}  }

\date{}
\maketitle \vskip 15mm
\begin{abstract}
We investigate the NLO QCD and the CP-violation effects in
$t\bar{t}Z^0$ production at the Large Hadron Collider(LHC) in the
minimal supersymmetric standard model(MSSM). Our calculation shows
that the total NLO QCD correction in the framework of the
CP-conserving MSSM significantly improves the scale uncertainty at
the leading order, and the contribution from the pure supersymmetric
QCD (pSQCD) correction can exceed $-8\%$ with the restrictions of
$90~GeV < p_T^t < 120~GeV$ and $120~GeV < p_T^Z < 150~GeV$, where
$p_T^t$ and $p_T^Z$ are the transverse momenta of the top-quark and
$Z^0$ gauge boson, respectively. Our numerical results demonstrate
that the pure supersymmetric QCD correction generally suppresses the total
SM-like QCD correction in the CP-conserving MSSM, and tends to be a
constant when either $\tilde{t}_1$ or $\tilde{g}$ is heavy enough.
We find also that the CP-odd asymmetry ${\cal A}_{\Phi}$ can reach
$2.17 \times 10^{-3}$, if the CP-phase angle really exists in the
coupling of gluino-stop-top.
\end{abstract}

\vskip 5mm

{\large\bf PACS: 12.60.Jv, 14.70.Hp, 14.65.Ha, 12.38.Bx } \vfill
\eject

\baselineskip=0.32in

\renewcommand{\theequation}{\arabic{section}.\arabic{equation}}
\renewcommand{\thesection}{\Roman{section}.}
\newcommand{\nb}{\nonumber}

\newcommand{\Dir}{\kern -6.4pt\Big{/}}
\newcommand{\Dirin}{\kern -10.4pt\Big{/}\kern 4.4pt}
\newcommand{\DDir}{\kern -7.6pt\Big{/}}
\newcommand{\DGir}{\kern -6.0pt\Big{/}}

\makeatletter      
\@addtoreset{equation}{section}
\makeatother       

\par
\section{Introduction}
\par
Although the standard model(SM)\cite{s1,s2} has achieved great
success in describing all the available experimental data, it
suffers from some conceptional difficulties. That has triggered an
intense activity in developing extension models. The
supersymmetric (SUSY) extensions\cite{Fayet:1976cr,
Nilles:1983ge,Haber:1984rc,Wess:1992cp,Yao:2006px}) rank among the
most promising and well-explored scenarios for new physics at the
TeV scale. Apart from predicting a light Higgs boson and
describing the low energy experimental data very well, the SUSY
models are able to solve various theoretical problems, e.g., the
SUSY models may provide an elegant way to construct the huge
hierarchy between the electroweak symmetry breaking(EWSB) scale
and the grand unification scale. At present the minimal
supersymmetric standard model(MSSM)\cite{a6} is regarded as the
simplest and the most attractive one in the SUSY models.

\par
The direct evidence for the top-quark was presented in 1995 by the
CDF and D0 collaborations of the Fermilab
Tevatron\cite{cdftop,d0top}. This is considered to be a remarkable
success of the SM. From that time on the top physics program has
been turned to precise investigation for its properties. Since the
top quark is the heaviest particle in the SM detected until now,
it plays a special role in the mechanism of the EWSB, and the new
physics connected to the EWSB may be found firstly through precise
study of top-quark observables. The high accumulated top-quark
events at the CERN LHC will open a new, rich field of top-quark
phenomenology. Deviations of experimental measurements from the SM
predictions, would indicate new non-standard top production or
decay mechanisms. Therefore, the precise study of the top
properties is one of the urgent priorities of the high energy
experimental program.

\par
Beside the SUSY particle direct production, virtual effects of
SUSY particles may induce deviations on observables from the SM
predictions. If SUSY particles are really detected at the LHC, the
comparison of precisely measured top-quark observables with the
theoretical predictions including SUSY loop effects may yield
additional information about the underlying model. Therefore,
probing precisely the properties of the top-quark is an important
goal at the LHC. In order to study precisely the top-quark physics
within the SM and beyond at the LHC, it is necessary to give the
theoretical predictions for top-quark observables including
higher-order corrections. In Ref.\cite{Berge} S. Berge et al.,
provide the predictions including the NLO SUSY QCD effects in the
MSSM for the total production rate and kinematic distributions of
polarized and unpolarized top-quark pair production at the
Tevatron and the LHC.

\par
Probing the couplings between the top quark and gauge bosons is
another way to discover new physics. Until now there have been
many works which devote to the observables related to the
top-quark gauge couplings in the SM and beyond. The theoretical
study of the effects of the top-quark and $Z^0$ gauge boson
coupling at colliders was widely carried out. The calculations for
the process $e^+e^- \to t\bar t Z^0$ at the leading-order(LO) and
including next-to-leading order(NLO) QCD, electroweak corrections
in the context of the SM were presented in Refs.
\cite{Hagiwara,DaiLei}, respectively, while CP-violating
effects in $e^+e^- \to t\bar t Z^0$ process were studied in the
framework of the two Higgs doublets model(THDM)\cite{CPT-THDM} and
with model independent effective Lagrangian\cite{CPT-Lag}. The
$\gamma\gamma \to t\bar t Z^0$ production channel has an
outstanding advantage over $e^+e^- \to t\bar t Z^0$ process in
measuring $t\bar tZ^0$ coupling at the ILC, due to its relatively
larger production rate\cite{chatterjee,denner0}. The NLO SUSY QCD
corrections to the $\gamma\gamma \to t\bar t Z^0$ process at
linear colliders, and the NLO SM QCD corrections to the $t\bar t
Z^0$ production at the LHC are studied in Refs.\cite{Dong} and
\cite{pp-ttz}, respectively.

\par
In this work, we calculate the production of the top-quark pair
associated with a $Z^0$ boson at the CERN LHC in both the
leading-order(LO) and NLO QCD approximations in the framework of
the MSSM with CP-conservation or CP-violation, and investigate the
possible CP-violating effects contributed by the CP-phase in the
couplings of gluino-stop-top predicted by the CP-violating MSSM.
The paper is organized as follows: The description of the related
theory about the CP-conserving and CP-violating MSSM is presented
in section 2. In section 3 we outline the technical details of the
related LO and NLO QCD calculations. In Sec.4 we give some
numerical results and discussions about the NLO SUSY QCD
corrections and the possible CP-odd effect. Finally, a short
summary is given.

\vskip5mm
\section{Related theory of the CP-violating MSSM }
\par
In the MSSM, each quark has two scalar partners called squarks,
$\tilde{q}_L$ and $\tilde{q}_R$(or $\tilde{q}_1$ and
$\tilde{q}_2$). The mass term of scalar quarks can be written
as\cite{s9}
\begin{eqnarray}\label{la-1}
 -{\cal L}_{\tilde{q}}^{mass} &=& \left(
\begin{array}{cc}
    \tilde{q}_L^{\dag} & \tilde{q}_R^{\dag}
\end{array}
    \right)
    {\cal M}^2_{\tilde{q}}     \left(
        \begin{array}{c}     \tilde{q}_{L} \\
       \tilde{q}_{R}
\end{array}
        \right),
\end{eqnarray}
where ${\cal M}^2_{\tilde{q}}$ is the mass squared matrix of
$\tilde q_L$ and $\tilde q_R$, expressed as
\begin{equation}\label{la-11}
{\cal M}^2_{\tilde{q}}= \left(
\begin{array}{cc}
m_{\tilde{q}_L}^2\ & a_qm_q
\\ a_q^*m_q\ & m_{\tilde{q}_R}^2 \\
\end{array}
\right).
\end{equation}
The diagonal and nondiagonal elements of this mass squared matrix
are
\begin{eqnarray}\label{sq-2}
m^2_{\tilde{q}_{L}}&=&\tilde{M}^2_{Q} + m^2_{q} +
      m_{Z}^2 (I_q^3- Q_q s_{W}^2) \cos{2 \beta}, \nb \\
m^2_{\tilde{q}_{R}}&=&\tilde{M}^2_{U,D} + m^2_{q} +
      Q_q m_{Z}^2 s_{W}^2 \cos{2 \beta},   \nb  \\
a_{q}m_{q}&=&m_{q}\left(A_q - \mu r_{U,D}\right),
\end{eqnarray}
where $m_q$, $Q_q$ and $I^3_q$ are the mass, electric charge and
the third component of the weak isospin of the quark $q$, $\mu$ is
the Higgsino mass parameter, $A_q$ ($q=u,d,c,s,t,b$) are the
supersymmetry soft-breaking trilinear coupling constants,
$\tilde{M}^2_Q$, $\tilde{M}^2_U$ and $\tilde{M}^2_D$ are the
supersymmetry soft-breaking mass parameters of the left- and
right-handed scalar quarks, and $r_U = 1/r_D = \cot\beta$ are for
the up- and down-type squarks, respectively. The mass squared
matrix ${\cal M}^2_{\tilde {q}}$ can be diagonalized by
introducing a unitary matrix ${\cal R}^{\tilde{q}}$. The mass
eigenstates $\tilde{q}_1$, $\tilde{q}_2$ are defined as
\begin{eqnarray}\label{la-3}
       \left(  \begin{array}{c} \tilde{q}_{1} \\
       \tilde{q}_{2} \end{array}
        \right) = {\cal R}^{\tilde q}  \left(
       \begin{array}{c}\tilde{q}_{L} \\
       \tilde{q}_{R} \end{array}
        \right).
\end{eqnarray}
Then the mass term of scalar quarks can be expressed as
\begin{eqnarray}\label{sq-0}
 -{\cal L}_{\tilde{q}}^{mass} &=&
 \left(
    \begin{array}{cc}
    \tilde{q}^{\dag}_{1} & \tilde{q}^{\dag}_{2}
     \end{array}
    \right)
    {\cal M}_D^{\tilde{q}~2 }
      \left(
        \begin{array}{c}
       \tilde{q}_{1} \\
       \tilde{q}_{2}
     \end{array}
        \right),
\end{eqnarray}
where
\begin{eqnarray}
{\cal M}_D^{\tilde{q}~2 } = {\cal R}^{\tilde q} {\cal M}^2_{\tilde
{q}} {\cal R}^{\tilde q~ \dag}=
   \left(
    \begin{array}{cc}
    m^2_{\tilde{q}_{1}} &  0 \\
    0 &  m^2_{\tilde{q}_{2}}
     \end{array}
    \right).
\end{eqnarray}
It is well known that the unitary matrix ${\cal R}^{\tilde{q}}$
can be parameterized as
\begin{eqnarray}\label{squark-matrix}
{\cal R}^{\tilde{q}} = \left(
\begin{array}{rr}
\cos \theta_q e^{-i \phi_q} & \sin \theta_q e^{i \phi_q} \\
-\sin \theta_q e^{-i \phi_q} & \cos \theta_q e^{i \phi_q}
\end{array}
\right),
\end{eqnarray}
where $\theta_q$ is called as the mixing angle between the left-
and right-handed squarks, and $2 \phi_q$ is the phase angle of
$a_q$ defined as $a_q = |a_q| e^{2 i \phi_q}$. The masses of the
squark mass eigenstates and the mixing angles acquire the forms as
\begin{eqnarray}
(m_{\tilde q_{1}}^2, m_{\tilde q_{2}}^2) &=& \frac{1}{2}\left\{
(m_{\tilde q_L}^2 + m_{\tilde q_R}^2 )\mp \left [ (m_{\tilde
q_L}^2 - m_{\tilde q_R}^2)^2 + 4|a_q|^2m_q^2 \right ]^{1/2} \right
\},  \nb \\
&& \tan{2\theta_{q}} =\frac{2
m_q|a_q|}{m^2_{\tilde{q}_{L}}-m^2_{\tilde{q}_{R}}}.
\end{eqnarray}
\par
Because of the large masses of the third generation quarks, the mixing
effects of the third generation squarks are more significant than
the first two generations. If we take the stop
masses($m_{\tilde{t}_1}$, $m_{\tilde{t}_2}$) and the stop mixing
angle($\theta_t$) as the input parameters for the stop sector, the
values of $m_{\tilde t_L}$, $m_{\tilde t_R}$ and $|a_t|$ can be
obtained by adopting Eq.(\ref{sq-1}).
\begin{eqnarray}\label{sq-1}
m^2_{\tilde{q}_{L}}&=&
      \cos^2\theta_{q}m^2_{\tilde{q}_1}+
      \sin^2\theta_{q}m^2_{\tilde{q}_2}, \nb \\
m^2_{\tilde{q}_{R}}&=&
      \sin^2\theta_{q}m^2_{\tilde{q}_1}+
      \cos^2\theta_{q}m^2_{\tilde{q}_2}, \nb \\
m_q |a_q|&=& \sin \theta_q \cos \theta_q \left(
m^2_{\tilde{q}_1}-m^2_{\tilde{q}_2} \right).
\end{eqnarray}

\par
In the CP-violating MSSM, the SUSY soft-breaking trilinear
coupling $A_{q}$ and the Higgsino mass parameter $\mu$ can be
complex. That makes $a_{q}$ having complex value. By using the
parameterization of the unitary matrix ${\cal R}^{\tilde{q}}$
(Eq.(\ref{squark-matrix})), we obtain the squark current
eigenstates ($\tilde{q}_L$, $\tilde{q}_R$) in terms of the mass
eigenstates ($\tilde{q}_1$, $\tilde{q}_2$) as
\begin{equation}\label{sq-8}
\tilde{q}_{L}=(\tilde{q}_{1}\cos\theta_{q}-
              \tilde{q}_{2}\sin\theta_{q})e^{i\phi_{q}},~~~~
\tilde{q}_{R}=(\tilde{q}_{1}\sin\theta_{q}+
               \tilde{q}_{2}\cos\theta_{q})e^{-i\phi_{q}}.
\end{equation}

\par
Normally the CP-violating effects in the MSSM from the
gluino-squark-quark interactions are much more important than from
the chargino and neutralino sectors due to the strong interaction.
We consider only the CP-violating effects induced by the
$\tilde{g}-\tilde{t}_{1,2}-t$ strong interactions. The Lagrangian
for the gluino-stop-top couplings is given by
\begin{eqnarray}\label{sq-9}
{\cal L}_{\tilde{g}-\tilde{t}-t} &=& \sqrt{2} g_s \sum_{a=1}^8
\bar{t} T^a \Big( \epsilon \tilde{t}_{R} P_L - \epsilon^{*}
\tilde{t}_{L} P_R \Big)
\tilde{g}^a + h.c. \nonumber \\
&=& \sqrt{2} g_s \sum_{a=1}^8 \sum_{\alpha, \beta = 1}^3
\bar{t}_{\alpha} T^{a}_{\alpha \beta} \Big( \epsilon \tilde{t}_{R
\beta} P_L - \epsilon^{*} \tilde{t}_{L \beta} P_R \Big)
\tilde{g}^a + h.c.,
\end{eqnarray}
where $g_s$ is the strong coupling constant, $T^a = (T^a_{\alpha
\beta})$ ($a=1,...,8$) are the $SU(3)$ generators, $a$, $\alpha$,
$\beta$ are the color indices of gluino, top and stop separately,
$P_{L,R} = (1 \mp \gamma^5)/2$ and $\epsilon = e^{-i
\phi_{SU(3)}}$. Here $2 \phi_{SU(3)}$ is the phase angle of
$M_{SU(3)}$, the supersymmetry soft-breaking $SU(3)$ gaugino
(gluino) mass parameter, which is defined as
\begin{eqnarray}
M_{SU(3)} = |M_{SU(3)}| e^{2 i \phi_{SU(3)}}.
\end{eqnarray}

\par
By inserting Eq.(\ref{sq-8}) into Eq.(\ref{sq-9}), the mixing
angle $\theta_{t}$ and phase angle $\phi_{t}$ may enter into the
couplings, and the Lagrangian is expressed in terms of the stop
mass eigenstates ($\tilde{t}_{1,2}$) instead of the current
eigenstates ($\tilde{t}_{L, R}$) as
\begin{eqnarray}
{\cal L}_{\tilde{g}-\tilde{t}-t} &=& \sqrt{2} g_s \sum_{a=1}^8
\bar{t} T^a \left[ \Big( \tilde{t}_1 \sin \theta_t + \tilde{t}_2
\cos \theta_t \Big) e^{-i \left( \phi_t + \phi_{SU(3)} \right)}
P_L \right. \\
&&\left. ~~~~~~~~~~~~~~~ - \Big( \tilde{t}_1 \cos \theta_t -
\tilde{t}_2 \sin \theta_t \Big) e^{i \left( \phi_t + \phi_{SU(3)}
\right)} P_R \right] \tilde{g}^a + h.c.~ . \nonumber
\end{eqnarray}
As shown in this Lagrangian, only the combination of the phase
angles $\phi_t$ and $\phi_{SU(3)}$ enters into the gluino-stop-top
couplings. Therefore, we redefine this combination as $\phi_t$,
\begin{eqnarray}
\phi_{t} + \phi_{SU(3)} \to \phi_t,
\end{eqnarray}
and obtain the conventional expression of the gluino-stop-top
interaction Lagrangian as
\begin{eqnarray}
{\cal L}_{\tilde{g}-\tilde{t}-t} &=& \sqrt{2} g_s \sum_{a=1}^8
\bar{t} T^a \left[ \Big( \tilde{t}_1 \sin \theta_t + \tilde{t}_2
\cos \theta_t \Big) e^{-i \phi_t} P_L
\right. \nonumber \\
&&\left. ~~~~~~~~~~~~~~~ - \Big( \tilde{t}_1 \cos \theta_t -
\tilde{t}_2 \sin \theta_t \Big) e^{i \phi_t} P_R \right]
\tilde{g}^a + h.c. \nonumber \\
&=& -i \sum_{i = 1}^2 \sum_{a=1}^8 \bar{t} \Big( V^L_{\tilde{g}
\tilde{t}_i t} P_L + V^R_{\tilde{g} \tilde{t}_i t} P_R \Big)
\tilde{g}^a \tilde{t}_i + h.c.,
\end{eqnarray}
where $V^L_{\tilde{g} \tilde{t}_1 t} = i \sqrt{2} g_s T^a \sin
\theta_t e^{-i \phi_t}$, $V^L_{\tilde{g} \tilde{t}_2 t} = i \sqrt{2}
g_s T^a \cos \theta_t e^{-i \phi_t}$,  $V^R_{\tilde{g} \tilde{t}_1
t} = -i \sqrt{2} g_s T^a \cos \theta_t e^{i \phi_t}$ and
$V^R_{\tilde{g} \tilde{t}_2 t} = i \sqrt{2} g_s T^a \sin \theta_t
e^{i \phi_t}$. There are similar expressions for other
$\tilde{g}-\tilde{q}_{i}-q$($q=u,d,c,s,b$) couplings involving
CP-phase angles. Because in this work we consider only the CP-phase
effects from the $\tilde{g}-\tilde{t}_{1,2}-t$ couplings, we take
$\phi_{t} \neq 0$ and $\phi_q = 0$ for $q = u,d,c,s,b$.

\par
In order to describe the CP-violating effects on the process, we
take a definition of a CP-odd observable for the LHC, which is
constructed to describe the distribution asymmetry of the
azimuthal angle $\Phi$ between $\hat{p}_{\bar{t}}^{T}$ and $\hat
{p}_{t}^{T}$ in the range of $-180^\circ \le \Phi \le 180^\circ$,
i.e.,
\begin{equation}\label{phi}
\Phi \equiv {\rm sgn}\left [(\vec{p}_t - \vec{p}_{\bar t})\cdot
\hat{z}\right ]\ {\rm sgn}\left [ (\vec{p}_t \times \vec{p}_{\bar
t})\cdot \hat{z}\right ]{\cos}^{-1} (\hat{p}_{t}^{T} \cdot
\hat{p}_{\bar{t}}^{T}),
\end{equation}
where $\Phi$ in Eq.(\ref{phi}) comes from the modified definition
of Eq.(14) in Ref.\cite{han}, and $\hat{z}$ is a unit vector of
the z-axis direction along one of the incoming proton. The
CP-asymmetry of angle $\Phi$ is defined as
\begin{equation}\label{CP-parameter}
{\cal A}_{\Phi} \equiv
\frac{\Delta\sigma_{\Phi}}{\sigma_T}=\frac{\sigma({180^\circ>\Phi>0^\circ})
- \sigma({0^\circ>\Phi>-180^\circ})
}{\sigma({180^\circ>\Phi>0^\circ})+\sigma({0^\circ>\Phi>-180^\circ})}.
\end{equation}
The significance is defined as
\begin{equation}\label{S-definition}
S= \frac{|\Delta\sigma_{\Phi}|{\cal L}}{\sqrt{\sigma_{T}{\cal L}}}.
\end{equation}
Then the CP-asymmetry effect may become observable at the $S\sigma$
significance, if the integrated luminosity has a value larger than
\begin{equation}\label{Luminosity}
{\cal L}=
S^2\frac{\sigma_T}{|\Delta\sigma_{\Phi}|^2}=\frac{S^2}{|{\cal
A}_{\Phi}|^2\sigma_T}.
\end{equation}

\vskip5mm
\par
\section{ Calculations }
\par
\subsection{ The LO cross sections for the partonic processes }
\par
\begin{figure}
\includegraphics[scale=1.0]{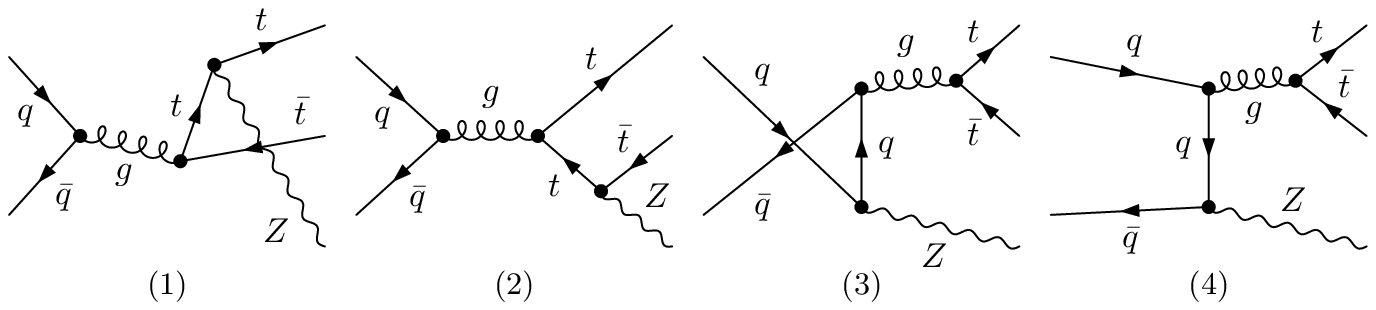}
\vspace*{0.3cm}\centering \caption{\label{fig1} The LO Feynman
diagrams for the \qqttz($q=u,d,s,c$) partonic process. }
\end{figure}
\begin{figure}
\includegraphics[scale=1.0]{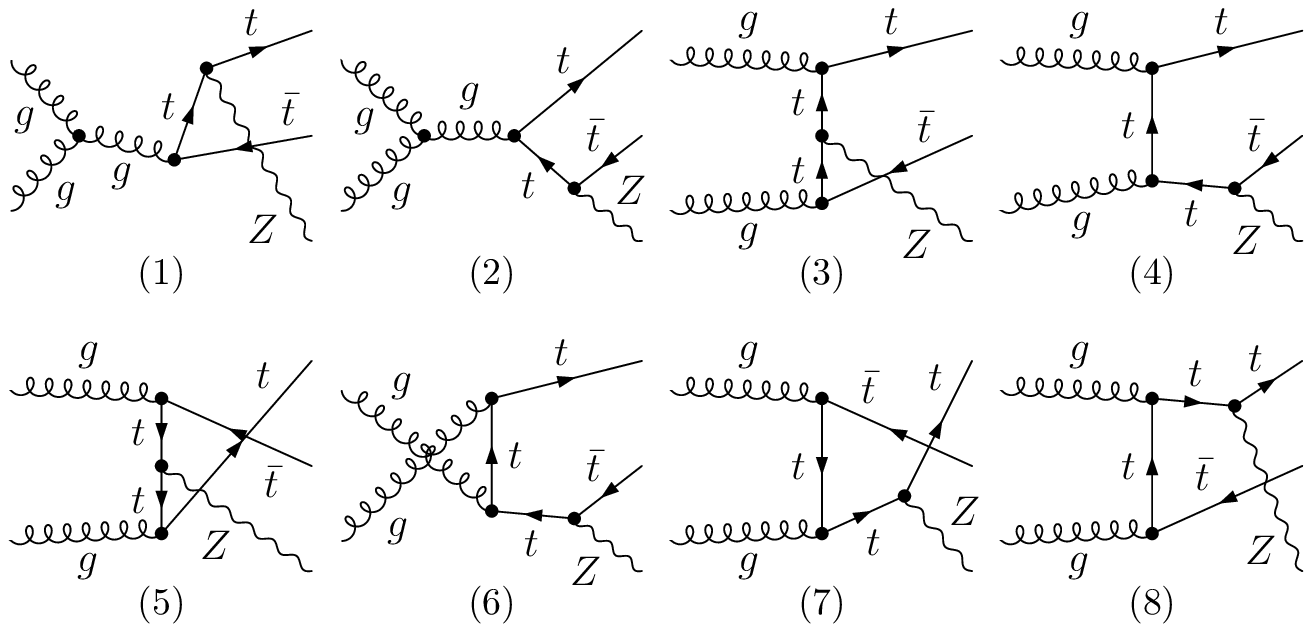}
\vspace*{0.3cm}\centering \caption{\label{fig2} The LO Feynman
diagrams for the \ggttz partonic process. }
\end{figure}

The contributions to the hadronic process of top-pair production
associated with a $Z^0$ boson at the LO, are from the partonic
processes \qqttz($q=u,d,c,s$) and \ggttz channels. We use the 't
Hooft-Feynman gauge in the following LO and NLO calculations. The LO
Feynman diagrams for the subprocesses $q(p_1)\bar q(p_2) \to
t(p_3)\bar t(p_4) Z^0(p_5)$, $(q=u,d,c,s)$ and $ g(p_1)g(p_2) \to
t(p_3)\bar t(p_4) Z^0(p_5)$ in the MSSM are depicted in
Fig.\ref{fig1} and Fig.\ref{fig2}, respectively.

\par
The LO Feynman diagrams for all these subprocesses in the MSSM are
the same as their corresponding ones in the SM. The explicit
expressions of the LO cross section for the partonic processes can
be written in the form as:
\begin{eqnarray}\label{int}
\hat{\sigma}_{LO}^{ij}= \frac{1}{4|\vec{p}_1|\sqrt{\hat{s}}}\int
{\rm d}\Gamma_3\overline{\sum} |{\cal M}_{LO}^{ij}|^2
\end{eqnarray}
where $ij=q\bar{q},gg$($q=u,d,c,s$), the summation is taken over the
spins and colors of initial and final states, $\vec{p}_1$ is the
c.m.s. momentum of one initial parton, and the bar over the
summation recalls averaging over the spins and colors of initial
partons. ${\rm d} \Gamma_3$ is the three-body phase space element
expressed as
\begin{eqnarray}
{\rm d} \Gamma_3 = (2\pi)^4  \delta^4 (p_1+p_2-\sum_{i = 3}^5 p_i)
\prod_{i=3}^5 \frac{d^3 \vec{p}_i}{(2\pi)^3 2E_i}.
\end{eqnarray}

\par
In the LO calculation for the parent process \ppttz we involve the
contributions from partonic processes \ggttz and \qqttz
($q=u,d,c,s$). Our numerical calculation shows the contribution to
the LO integrated cross section from the partonic processes $s\bar
s, c\bar c \to t\bar tZ^0$ is less than $3\%$ at the LHC. Therefore,
we consider only the NLO QCD corrections to the processes $pp \to
u\bar u, d\bar d, gg \to t\bar tZ^0+X$ in the following NLO
calculation.

\par
\subsection{ NLO QCD corrections to the partonic processes }
\par
The NLO QCD correction in the MSSM(NLO SQCD) to each of the
partonic subprocess \qqttz ($q=u,d$) and \ggttz consists of two
independent parts. One is the so-called SM-like component, another
is the pure SUSY QCD (pSQCD) component arising from the
contributions of the virtual gluino one-loop diagrams. We adopt
the dimensional regularization scheme in $D = 4-2\epsilon$
dimensions to isolate the ultraviolet (UV) and infrared (IR)
singularities. Then the total NLO SQCD corrections to the partonic
subprocess \qqttz($q=u,d$) and \ggttz, can be written as:
\begin{eqnarray}
\Delta\hat{\sigma}_{SNLO}^{ij} = \Delta\hat{\sigma}_{SM-like}^{ij}
+ \Delta\hat{\sigma}_{pSQCD}^{ij},~~~(ij=u\bar u,d \bar d,gg).
\end{eqnarray}

\par
In the MSSM, the so-called SM-like NLO QCD correction component is
exactly equal to the NLO QCD correction in the SM, and we shall
compare our results for the SM-like correction with those in
Ref.\cite{pp-ttz}. The pSQCD correction component is UV and IR
finite after renormalzation. The NLO SQCD correction includes the
following contributions:
\par
$\blacktriangleright$ the virtual corrections to the partonic
process \qqggttz.
\par
$\blacktriangleright$ the real gluon emission partonic process
\qqggttzg.
\par
$\blacktriangleright$ the real light-(anti)quark emission partonic
process \qgttzq.
\par
$\blacktriangleright$ the collinear counterterms of the PDF.

\par
{\bf (1) Virtual corrections in the MSSM}
\par
\begin{figure}
\begin{center}
\includegraphics[scale=1.2]{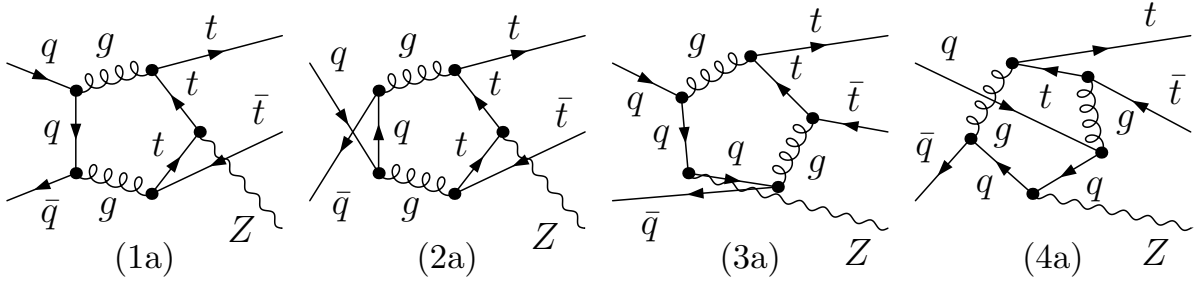}
\caption{\label{fig3} The SM-like pentagon Feynman diagrams for
the \qqttz partonic process. }
\end{center}
\end{figure}
\begin{figure}
\begin{center}
\includegraphics[scale=1.2]{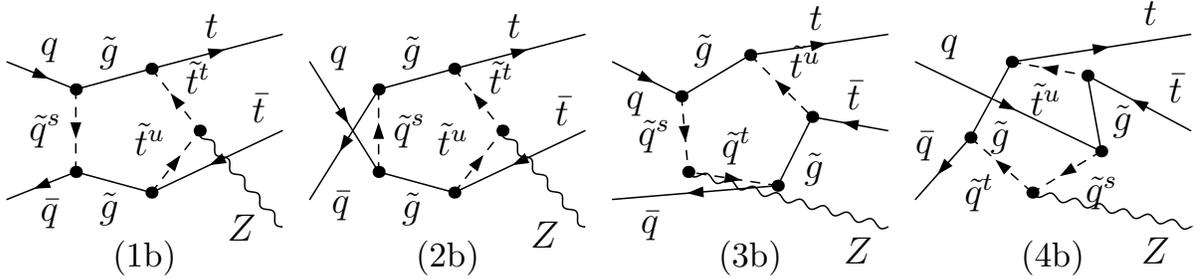}
\caption{\label{fig4} The pSQCD pentagon Feynman diagrams for the
\qqttz partonic process with the upper indices in
$\tilde{t}^{u,t}$ and $\tilde{q}^{s,t}$ running from 1 to 2
respectively. }
\end{center}
\end{figure}
\begin{figure}
\begin{center}
\includegraphics[scale=1.2]{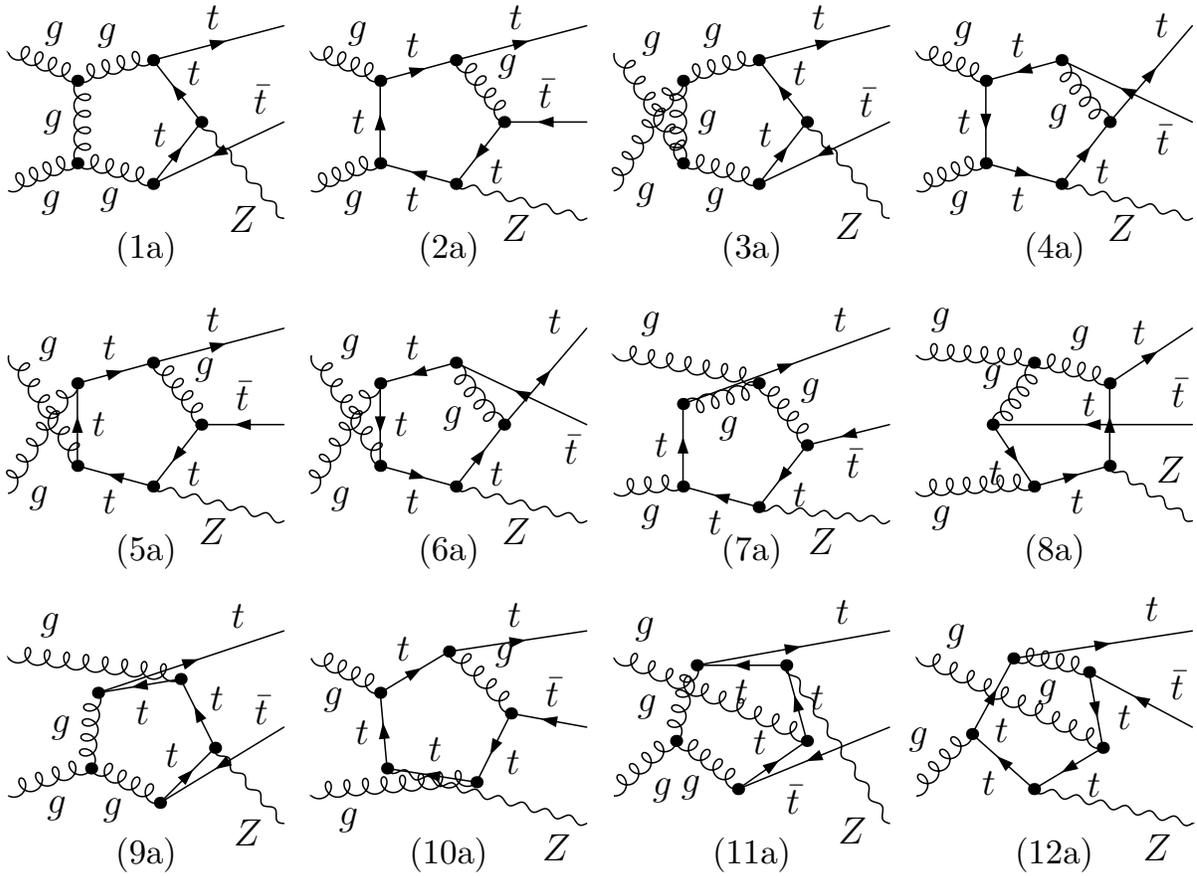}
\caption{\label{fig5} The SM-like pentagon Feynman diagrams for the
\ggttz partonic process. The diagrams obtained by exchanging initial
gluons are not depicted. }
\end{center}
\end{figure}
\begin{figure}
\begin{center}
\includegraphics[scale=1.2]{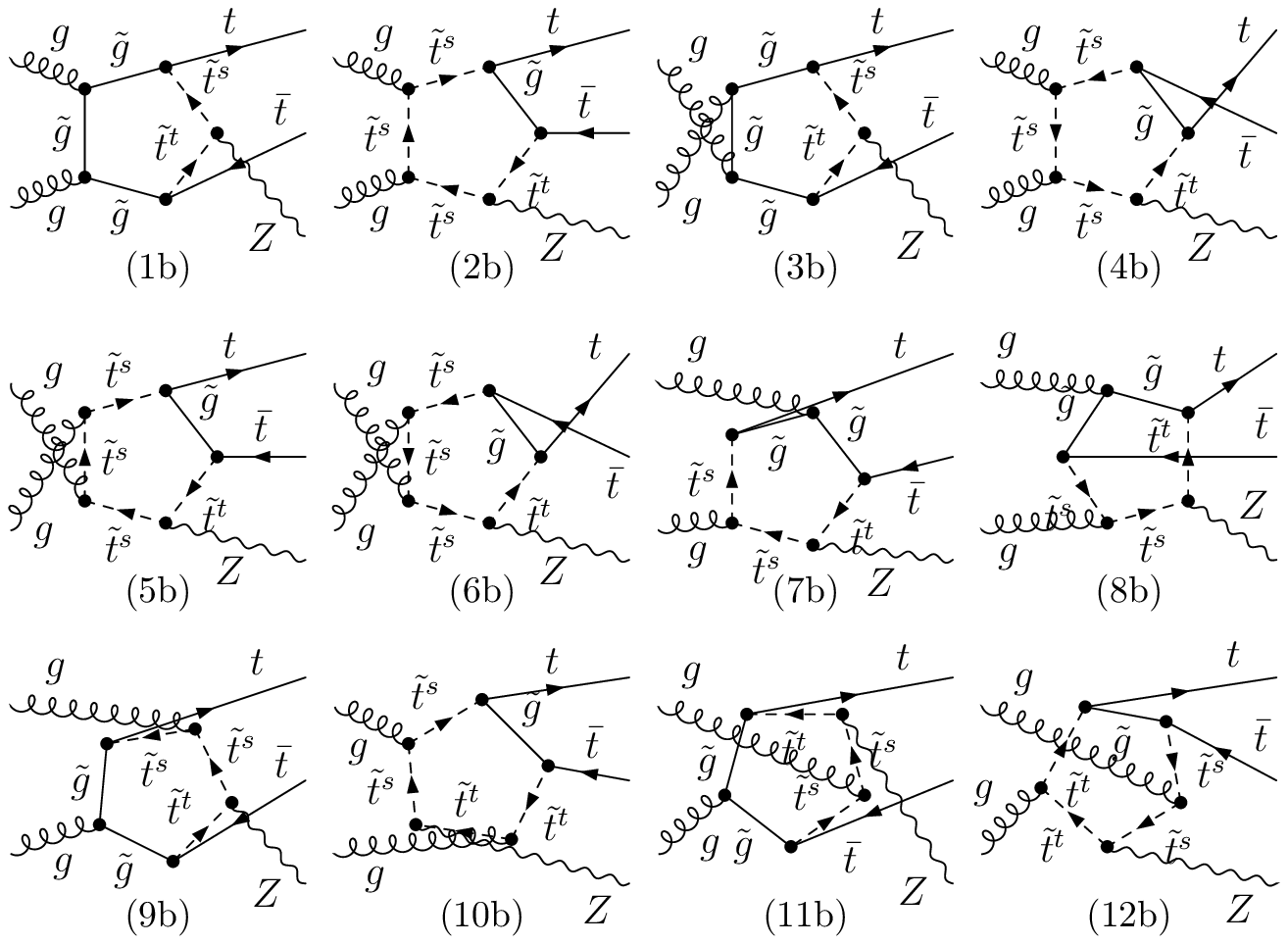}
\caption{\label{fig6} The pSQCD pentagon Feynman diagrams for the
\ggttz partonic process. The upper indices in $\tilde{t}^{s,t}$
run from 1 to 2 respectively. The diagrams obtained by exchanging
initial gluons are not shown. }
\end{center}
\end{figure}

In the MSSM, the virtual QCD ${\cal O} ({\alpha_s})$ corrections
come from the one-loop diagrams including self-energy, vertex, box
and pentagon diagrams. In Figs.\ref{fig3}-\ref{fig6}, we
illustrate all the pentagon graphs for the partonic processes \qqttz
and \ggttz, separately. We take the definitions of the scalar and
tensor two-, three-, four- and five-point integral functions
presented in Ref.\cite{denner}. We use Passarino-Veltman
method\cite{Passar} to reduce the N-point($N\leq 5$) tensor
functions to scalar integrals, and manipulate the $\gamma_5$ matrix
in D-dimensions by employing a naive scheme as presented in
Ref.\cite{Chanowitz}, which keeps an anticommuting $\gamma_5$ in all
dimensions. The one-loop Feynman diagrams and the corresponding
amplitudes are created by using FeynArts3.2 package\cite{Hahn}, and
the scalar integrals are evaluate mainly by adopting the
LoopTools-2.1 package\cite{formloop,ff}. In order to cancel the UV
divergences from both the SM-like and pSQCD one-loop diagrams, we
should introduce some suitable counterterms.
\begin{eqnarray}
m_t & \to & m_t+\delta m_t=m_t+\delta m_{t}^{SM-like}+\delta m_{t}^{pSQCD},    \nb \\
g_s &\to& g_s+\delta g_s=g_s+\delta g_s^{SM-like}+\delta g_s^{pSQCD},    \nb \\
t_L  & \to & (1+\frac{1}{2}\delta Z_{L}^t) t_L=\left [
1+\frac{1}{2}(\delta Z_{L}^{t,SM-like}+\delta Z_{L}^{t,pSQCD})
\right
]t_L,    \nb  \\
 t_R & \to & (1+\frac{1}{2}\delta Z_{R}^t) t_R=\left [1+\frac{1}{2}(\delta
Z_{R}^{t,SM-like}+\delta Z_{R}^{t,pSQCD}) \right]t_R ,   \nb  \\
G_{\mu}^a & \to & (1+ \frac{1}{2}\delta Z_g) G_{\mu}= \left[1+
\frac{1}{2}\left(\delta Z_g^{SM-like}+ \delta Z_g^{pSQCD}\right)
\right ] G_{\mu}^a,
\end{eqnarray}
where $t_{L,R}$ and $G_{\mu}$ are the wave functions of top-quark
and gluon, respectively.

\par
The counterterms of top-quark, gluon fields and top-quark mass are
fixed by using on-mass-shell renormalization conditions
\cite{denner}. For the renomalization of the QCD strong coupling
constant $g_s$, we use the $\overline{MS}$ scheme except that the
divergences associated with top quark and the colored SUSY
particle loops are subtracted at zero momentum\cite{gs}. The
counterterm of the strong coupling constant includes the SM-like
and pSQCD terms, which can be expressed as follows,
\begin{eqnarray}
\frac{\delta g_s^{SM-like}} {g_s} = - \frac {\alpha_s(\mu_r)}{4 \pi}
\left [\frac{\beta_0^{SM-like}} {2} \frac{1}{\bar{\epsilon}} +
\frac{1}{3} \ln\frac{m_t^2}{\mu_r^2}\right ],
\end{eqnarray}

\begin{eqnarray}
&& \frac{\delta g_s^{pSQCD}} {g_s} = - \frac {\alpha_s(\mu_r)}{4
\pi} \left[\frac{\beta_1^{pSQCD}}{2\bar{\epsilon}} +
\frac{N_c}{3}\ln\frac{m_{\tilde{g}}^2}{\mu_r^2} +
\sum_{U=u,c,t}^{i =
1,2}\frac{1}{12}\ln\frac{m_{\tilde{U_i}}^2}{\mu_r^2} +
\sum_{D=d,s,b}^{j =
1,2}\frac{1}{12}\ln\frac{m_{\tilde{D_j}}^2}{\mu_r^2}\right], \nb
\\
\end{eqnarray}
where
\begin{eqnarray}
\beta_0^{(SM-like)} = \frac{11}{3} N_c - \frac{2}{3} n_{lf} -
\frac{2}{3}, ~~~\beta_1^{(pSQCD)} = -\frac{2}{3} N_c - \frac{1}{3}
(n_{lf} + 1),
\end{eqnarray}
with $N_c = 3$, $n_{lf} = 5$ light flavors and
$\frac{1}{\bar{\epsilon}} = \frac{1}{\epsilon_{UV}} + \ln(4 \pi) -
\gamma_E$. With the introduction of the CP-violating phase
$\phi_t$ in the MSSM, the renormalized one-particle irreducible
two-point functions for top-quark and gluon containing the
contributions from pSQCD are defined as follows
\cite{denner,Bernd}
\begin{eqnarray}
\hat{\Gamma}^{(t)}_{pSQCD}(p) =  i \left [ \rlap/p P_{L}
     \hat{\Sigma}^{(t)L}_{pSQCD}(p^2) + \rlap/p P_{R} \hat{\Sigma}^{(t)R}_{pSQCD}(p^2)
    + P_{L} \hat{\Sigma}^{(t)S}_{pSQCD}(p^2) + P_{R}
    \hat{\Sigma}^{(t)S~\ddag}_{pSQCD}(p^2) \right],  \nb
\end{eqnarray}
\begin{eqnarray}\label{self-energy}
\hat{\Gamma}^{(g)ab}_{pSQCD}(p) = -i \left(g^{\mu\nu}-
\frac{p^{\mu}p^{\nu}}{p^2} \right)
\delta^{ab}\hat{\Sigma}^{(g)T}_{pSQCD}(p^2)-i\frac{p^{\mu}p^{\nu}}{p^2}
\delta^{ab}\hat{\Sigma}^{(g)L}_{pSQCD}(p^2).
\end{eqnarray}
It should be mentioned here that in the first equation of
Eqs.(\ref{self-energy}) the upper conjugation symbol $\ddag$ acts
only on the CP-violating phase. The SM-like components for the top
quark, gluon self-energies, wave function and top mass
counterterms, can be found in many references, such as
Ref.\cite{{denner}}. Here we present only the related pSQCD
component expressions for unrenormalized top quark, gluon
self-energies and counterterms.
\begin{eqnarray}
\Sigma^{(t)L}_{pSQCD}(p^2) = -\frac{C_F}{8 \pi^2} g_{s}^2
 \left ( \cos^2\theta_{\tilde{t}}~ B_{1}[p^2, m_{\tilde{g}}^2, m_{\tilde{t}_1}^2]
 + \sin^2\theta_{\tilde{t}}~ B_{1}[p^2, m_{\tilde{g}}^2, m_{\tilde{t}_2}^2] \right ),
\end{eqnarray}
\begin{eqnarray}
\Sigma^{(t)R}_{pSQCD}(p^2) = -\frac{C_F}{8 \pi^2} g_{s}^2
        \left ( \sin^2\theta_{\tilde{t}}~B_{1}[p^2, m_{\tilde{g}}^2, m_{\tilde{t}_1}^2]
         + \cos^2\theta_{\tilde{t}}~B_{1}[p^2, m_{\tilde{g}}^2, m_{\tilde{t}_2}^2] \right ),
\end{eqnarray}
\begin{eqnarray}
\Sigma^{(t)S}_{pSQCD}(p^2) = -\frac{C_F}{8 \pi^2} g_{s}^2
m_{\tilde{g}} \left(
\sin\theta_{\tilde{t}}~\cos\theta_{\tilde{t}}~e^{-2i\phi}\right)
\left ( B_{0}[p^2, m_{\tilde{g}}^2, m_{\tilde{t}_1}^2]
 - B_{0}[p^2, m_{\tilde{g}}^2, m_{\tilde{t}_2}^2] \right ),
\end{eqnarray}
\begin{eqnarray}
\Sigma^{(t)S~\ddag}_{pSQCD}(p^2) = -\frac{C_F}{8 \pi^2} g_{s}^2
m_{\tilde{g}}
\left(\sin\theta_{\tilde{t}}~\cos\theta_{\tilde{t}}~e^{2i\phi}\right)
\left ( B_{0}[p^2, m_{\tilde{g}}^2, m_{\tilde{t}_1}^2]
 - B_{0}[p^2, m_{\tilde{g}}^2, m_{\tilde{t}_2}^2] \right ),
\end{eqnarray}

\par
Since we don't need the longitudinal part of gluon self-energy in
the following calculation, we give the explicit expression for
$\Sigma^{(g)T}_{pSQCD}(p^2)$ only.
\begin{eqnarray}
\Sigma^{(g)T}_{pSQCD}(p^2) &=& \frac{g_s^2}{16 \pi^2} \left \{
  3D \left [ (2-D)B_{00}+ m^2_{\tilde{g}}B_0-p^2(B_{11}+B_1)\right]
[p^2,m^2_{\tilde{g}},m^2_{\tilde{g}}] \right. \nb \\
&+& \left. \sum_{i=1}^{2}\sum_{q=u,d,c}^{s,t,b} \left
(A_0[m^2_{\tilde{q}_i}]-2B_{00}[p^2,m^2_{\tilde{q}_i},m^2_{\tilde{q}_i}]
\right) \right \},
\end{eqnarray}
where $D = 4-2\epsilon$ and the definitions of the two-point
integrals are adopted from Ref.\cite{denner}. In the $SU(3)$ group,
$C_F = (N_c^2 - 1)/(2 N_c)$. By using the relevant on-mass-shell
renormalization conditions and imposing the real condition on the
right-handed top-quark field renormalization constant, $\delta Z_R^t
= \delta Z_R^{t \dag}$\cite{HanL}, we obtain
\begin{eqnarray}\label{eq7}
\delta m_{t}^{pSQCD} = \frac{1}{2}
        \left ( m_t \tilde{Re} \Sigma^{(t)L}_{pSQCD}(m_{t}^{2}) +
                m_t \tilde{Re} \Sigma^{(t)R}_{pSQCD}(m_{t}^{2}) +
                \tilde{Re} \Sigma^{(t)S}_{pSQCD}(m_{t}^{2}) +
                \tilde{Re} \Sigma^{(t)S~\ddag}_{pSQCD}(m_{t}^{2}) \right
                ),  \nb \\
\end{eqnarray}
\begin{eqnarray}\label{eq8}
\delta Z_{L}^{t,pSQCD} &=& -\Sigma^{(t)L}_{pSQCD}(m_{t}^{2}) -
    \frac{1}{m_t} \left [ \tilde{Re} \Sigma^{(t)S~\ddag}_{pSQCD}(m_t^2) -
                          \tilde{Re} \Sigma^{(t)S}_{pSQCD}(m_t^2) \right ] \nb \\
    &-& m_{t} \frac{\partial}{\partial p^2}
        \left [ m_t \tilde{Re} \Sigma^{(t)L}_{pSQCD}(p^2) +
                m_t \tilde{Re}\Sigma^{(t)R}_{pSQCD}(p^2)   \right.   \nb \\
    &+& \left.        \tilde{Re} \Sigma^{(t)S}_{pSQCD}(p^2) +
                \tilde{Re} \Sigma^{(t)S~\ddag}_{pSQCD}(p^2) \right ]
                |_{p^2=m_{t}^2},  \nb \\
\end{eqnarray}
\begin{eqnarray}\label{eq9}
\delta Z_{R}^{t,pSQCD} &=& -\Sigma^{(t)R}_{pSQCD}(m_t^2) -
    m_{t} \frac{\partial}{\partial p^2}
        \tilde{Re}\left [ m_t  \Sigma^{(t)L}_{pSQCD}(p^2) +
         m_t \Sigma^{(t)R}_{pSQCD}(p^2) \right.   \nb \\
    &+& \left .   \Sigma^{(t)S}_{pSQCD}(p^2)  +
                \Sigma^{(t)S~\ddag}_{pSQCD}(p^2) \right ]
                |_{p^2=m_{t}^2},
\end{eqnarray}
\begin{eqnarray}\label{eq10}
\delta Z_g^{pSQCD} = -\tilde{Re}\frac{\partial
\Sigma^{(g)T}_{pSQCD} (p^2)}{\partial p^2}|_{p^2=0},
\end{eqnarray}
where $\tilde{Re}$ only takes the real part of the loop integral
functions appearing in the self-energies. The renormalized
amplitudes of all the NLO QCD virtual corrections to the partonic
processes \qqttz and \ggttz in the MSSM are expressed as
\begin{eqnarray}\label{eq11}
\Delta {\cal M}_{vir}^{ij} =  \Delta {\cal M}_{self}^{ij} + \Delta
{\cal M}_{tri}^{ij} + \Delta{\cal M}_{box}^{ij} + \Delta{\cal
M}_{pent}^{ij}+ \Delta{\cal M}_{count}^{ij}, ~~~(ij=u\bar u,d\bar
d,gg),
\end{eqnarray}
where $\Delta{\cal M}_{self}^{ij}$, $\Delta{\cal M}_{tri}^{ij}$,
$\Delta{\cal M}_{box}^{ij}$, $\Delta{\cal M}_{pent}^{ij}$ and
$\Delta{\cal M}_{count}^{ij}$ represent the amplitudes for
self-energy, triangle, box, pentagon and counterterm diagrams,
respectively. Then we can get the UV-finite virtual NLO QCD
correction component $\Delta\hat{\sigma}_{vir}^{ij}$ as
\begin{eqnarray}
\Delta\hat{\sigma}_{vir}^{ij}=
\frac{1}{2|\vec{p}_1|\sqrt{\hat{s}}} \int {\rm
d}\Gamma_3\overline{\sum} {\rm {Re}}( {\cal M}_{LO}^{ij} \times
\Delta{\cal M}_{vir}^{ij} ).
\end{eqnarray}
The definitions of the notations appeared in above equation are
the same with those in Eq.(\ref{int}).

\par
{\bf (2) Real gluon and light-(anti)quark emission corrections }
\par
In the MSSM, the real ${\cal O}(\alpha_s)$ correction processes
involve the real gluon emission and real light-(anti)quark
emission processes which are listed as follows:
\begin{eqnarray}
q(p_1) + \bar{q}(p_2) \rightarrow t(p_3) + \bar{t}(p_4) + Z^0(p_5) +
g(p_6)
\end{eqnarray}
\begin{eqnarray}
g(p_1) + g(p_2) \rightarrow t(p_3) + \bar{t}(p_4) + Z^0(p_5) +
g(p_6)
\end{eqnarray}
\begin{eqnarray}
q(\bar{q})(p_1) + g(p_2) \rightarrow t(p_3) +\bar{t}(p_4) + Z^0(p_5)
+ q(\bar{q})(p_6).
\end{eqnarray}
Because of  the IR singularities involved in these processes, we use
the two cutoff phase space slicing method (TCPSS) to perform the
integration over the phase space of these real emission
processes.\cite{TCPSS}. In our calculations, the real gluon
emission correction to each of the processes $ij \to t\bar t
Z^0,~~(ij=u\bar u,d\bar d,gg)$ contains both soft and collinear IR
singularities, which are involved in soft gluon region($E_6 \leq
\delta_s \sqrt{\hat s} /2$) and hard gluon region($E_6 > \delta_s
\sqrt{\hat s} /2$) respectively. The hard gluon region is also
divided into the hard collinear region (HC) and the hard
noncollinear region ($\overline{HC}$) with $\frac {2 p_i .
p_6}{E_6 \sqrt{\hat s}} < \delta_c$ and $\frac {2 p_i . p_6}{E_6
\sqrt{\hat s}} \geq \delta_c$ ($p_i$ are the momenta for $q$ and
$\bar q$). Each of the real light-(anti)quark emission processes
contains only collinear IR singularity, and can be dealt with in
the hard collinear region(HC) too. In the $\overline{HC}$ region,
the real emission corrections, $\Delta\hat
\sigma_{\overline{HC}}^{kl}$, where $kl=q \bar{q}, gg, qg,
\bar{q}g$, $(q=u,d)$, are finite and can be calculated numerically
with general Monte Carlo method. After summing the virtual and
real gluon/(anti)quark radiation corrections, the remained
collinear divergence can be cancelled by that in the NLO PDFs.
Then the finite total NLO QCD correction to the \ppttz process can
be obtained.

\par
\subsection{Total cross sections for the \ppttz process}
\par
The LO, NLO SQCD corrected hadronic cross sections for \ppttz in
the MSSM can be written as:
\begin{eqnarray}\label{total-cross-section}
d\sigma_{LO,SNLO}(pp \to t\bar{t}Z^0+X ) &=& \sum_{ij=u\bar
u,d\bar d}^{c\bar c,s\bar s,gg}
\frac{1}{1+\delta_{ij}}\int_0^1 dx_1 dx_2 \nonumber \\
&& \times [G_i(x_1,\mu_f)
G_j(x_2,\mu_f)d\hat\sigma_{LO,SNLO}^{ij}(x_1x_2\sqrt{s},\mu_r) +(1
\leftrightarrow 2)].
\end{eqnarray}
We adopt the CTEQ6L1 and CTEQ6m PDFs\cite{cteq} in the LO and NLO
calculations respectively, except in the specific case for
numerical comparison. The $\hat\sigma_{LO,SNLO}^{ij}(x_1
x_2\sqrt{s},\mu_r)$ are the LO, NLO SQCD corrected cross sections
with the partonic colliding energy $\sqrt{\hat{s}}=x_1x_2\sqrt{s}$
for the partonic processes of $ij \to t\bar{t}Z^0$($ij=gg, q\bar
q$) in the MSSM. Throughout our evaluation, we equate the
factorization and renormalization scales and define
$\mu=\mu_f=\mu_r$.

\par
In the MSSM, the NLO SQCD corrected partonic cross sections can be
expressed as below:
\begin{eqnarray}
d\hat\sigma^{ij}_{SNLO}(x_1 x_2\sqrt{s},\mu) =
d\hat\sigma^{ij}_{LO}(x_1 x_2\sqrt{s},\mu) +
d\Delta\hat\sigma^{ij}_{SNLO}(x_1 x_2\sqrt{s},\mu)
\end{eqnarray}
where $ij=gg, q\bar q$, $\Delta\hat\sigma^{ij}_{SNLO}(x_1,
x_2,\mu)$ denotes the total NLO QCD correction in the MSSM to the
corresponding LO partonic cross section. In this work we ignore
reasonably the NLO QCD corrections to the partonic processes
$c\bar c,s\bar s \to t\bar t Z^0$ due to the luminosity
suppression, i.e., $\Delta\hat\sigma^{c\bar c,~s\bar s}_{SNLO}(x_1
x_2\sqrt{s},\mu)=0$. Then the full NLO QCD corrections in the MSSM
to the process \ppttz at the LHC can be divided as:
\begin{eqnarray}
\Delta\sigma_{SNLO} = \Delta\sigma_{SM-like} +
\Delta\sigma_{pSQCD}.
\end{eqnarray}
The later part arises from the virtual correction of the diagrams
with gluino/squark in loops. Note that the cross sections for the
parent process \ppttz $\sigma_{SM-like}$ should be the same as the
corresponding ones, $\sigma_{SM}$, in the SM\cite{pp-ttz,ttz3}.

\par
\section{Numerical Results and discussions}
\par
In this section, we present numerical results of the NLO QCD
corrections to the process \ppttz in the MSSM at the LHC. The
numerical results for the LO and the NLO SM-like QCD corrections
have been compared with the data presented in Tabel 1 of
Ref.\cite{ttz3}. Both result sets are listed in Table \ref{tab1}.
There we employ the MRST PDFs\cite{mrst} and the input parameters
which were used in Ref.\cite{ttz3}. The agreement between them can
be seen obviously from the table.
\begin{table}
\begin{center}
\begin{tabular}{|c|c|c|c|c|}
\hline $\mu $ & our $\sigma_{LO}$(pb) & our $\sigma_{NLO}$(pb) &
$\sigma_{LO}$(pb) in Ref.\cite{ttz3} & $\sigma_{NLO}$(pb) in Ref.\cite{ttz3}   \\
\hline   $\mu_0/4$   & 1.0779(8) & 1.216(5)  &  1.078  & 1.213       \\
\hline   $\mu_0/2$   & 0.8083(6) & 1.095(4)  &  0.808  & 1.093       \\
\hline   $\mu_0$     & 0.6198(3) & 0.975(4)  &  0.620  & 0.973       \\
\hline
\end{tabular}
\end{center}
\begin{center}
\begin{minipage}{15cm}
\caption{\label{tab1} The comparison of our numerical results for
the \ppttz process in the SM at the LHC with those in
Ref.\cite{ttz3}, where the LO and NLO QCD corrected cross sections
for different energy scale values($\mu=\mu_f=\mu_r$) are listed
with the relevant parameters and the PDFs being the same as in
Ref.\cite{ttz3}, i.e., $\mu_0 = 2 m_t + m_Z$, $m_t = 170.9~GeV$,
$m_Z = 91.19~GeV$, $m_W = 80.45~GeV$ and the MRST PDFs. Note that
the numerical results are contributed by the subprocesses $q\bar
q,gg \to t\bar t Z^0$ with $q=u,d,c,s$ at the LO, but $q=u,d$ at
the NLO. }
\end{minipage}
\end{center}
\end{table}

\par
In the following numerical calculations in the frameworks of the
SM and MSSM, we define $\mu_0 \equiv m_t + m_Z/2$ and take CTEQ6L1
PDFs with an one-loop running $\alpha_s$ in the LO calculation and
CTEQ6M PDFs with a two-loop $\alpha_s$ in the NLO
calculation\cite{hepdata}. The number of active flavors is $N_f=5$
and the QCD parameters are $\Lambda_5^{LO}=166~MeV$ and
$\Lambda_5^{\overline{MS}}=227~MeV$ for the LO and NLO
calculations, respectively\cite{hepdata}. We ignore the masses of
$u$-, $d$-, and $s$-quarks in our calculations. The other SM
parameters are taken as\cite{hepdata},
\begin{eqnarray}\label{SM-parameter}
\alpha_{ew}({m_Z}^2) = 1/127.918,~~m_t = 171.2~GeV,
~~m_b = 4.2~GeV,~~m_c=1.3~GeV,  \nb \\
m_Z = 91.1876~GeV,~~m_W =
80.398~GeV,~~\sin^2\theta_W=1-m_W^2/m_Z^2=0.222646.
\end{eqnarray}

\par
For the SUSY parameters of the scalar top sector in the
CP-conserving MSSM, we take the top squark
masses($m_{\tilde{t}_1}$, $m_{\tilde{t}_2}$) and their mixing
angle($\theta_{t}$) as input parameters, and adopt Eq.(\ref{sq-1})
to calculate the $m^2_{\tilde{t}_L}$, $m^2_{\tilde{t}_R}$, $a_t$,
and sequentially use Eq.(\ref{sq-2}) to get the SUSY soft-breaking
parameters $\tilde{M}_{Q_t}$ and $\tilde{M}_{U_t}$ for the stop
sector by assuming $\tan\beta=10$(Here we take $\tan \beta=10$
arbitrarily, because the NLO SQCD cross sections for $pp \to t\bar
tZ^0+X$ do not directly related to $\tan\beta$, and in the range
of $1<\tan \beta<50$ the dependences of $m_{\tilde{t}_1}$ and
$m_{\tilde{t}_2}$ on $\tan \beta$ are rather weak\cite{Berge}).
Since the gauge invariance in the MSSM requires $\tilde{M}_{Q_b} =
\tilde{M}_{Q_t}$, we need only to fix two additional parameters
for the scalar bottom sector. We neglect the mixing for the scalar
bottom sector and assume $m_{\tilde{b}_L}=m_{\tilde{b}_R}$, the
masses of $\tilde{b}_{1,2}$ can be obtained by the following
equation,
\begin{eqnarray}\label{mass-sbottom}
m^2_{\tilde{b}_{L}}&=&\tilde{M}^2_{Q_t} + m^2_{b} +
      m_{Z}^2 (I_b^3- Q_b s_{W}^2) \cos{2 \beta}.
\end{eqnarray}

\par
Furthermore, we take $m_{\tilde{g}}=200~GeV$\cite{para}, $\phi_q=0$
(for the u-, d-, c-, s-, b-quarks) and let CP-phase
$\phi\equiv\phi_t\neq0$ as a free parameter in the CP-violating
MSSM. For the first two squark generations, we take the assumption
of the universal squark mass as used in Ref.\cite{Berge} i.e.,
\begin{eqnarray}\label{mass-squark}
m_{\tilde{u}_{L}}=m_{\tilde{u}_{R}}=m_{\tilde{d}_{L}}=m_{\tilde{d}_{R}}=
m_{\tilde{c}_{L}}=m_{\tilde{c}_{R}}=m_{\tilde{s}_{L}}=m_{\tilde{s}_{R}}=
m_{\tilde{q}},
\end{eqnarray}
and set $m_{\tilde{q}}=1~TeV$ in the following calculations.

\par
\begin{figure}[htbp]
\includegraphics[width=3.2in,height=3in]{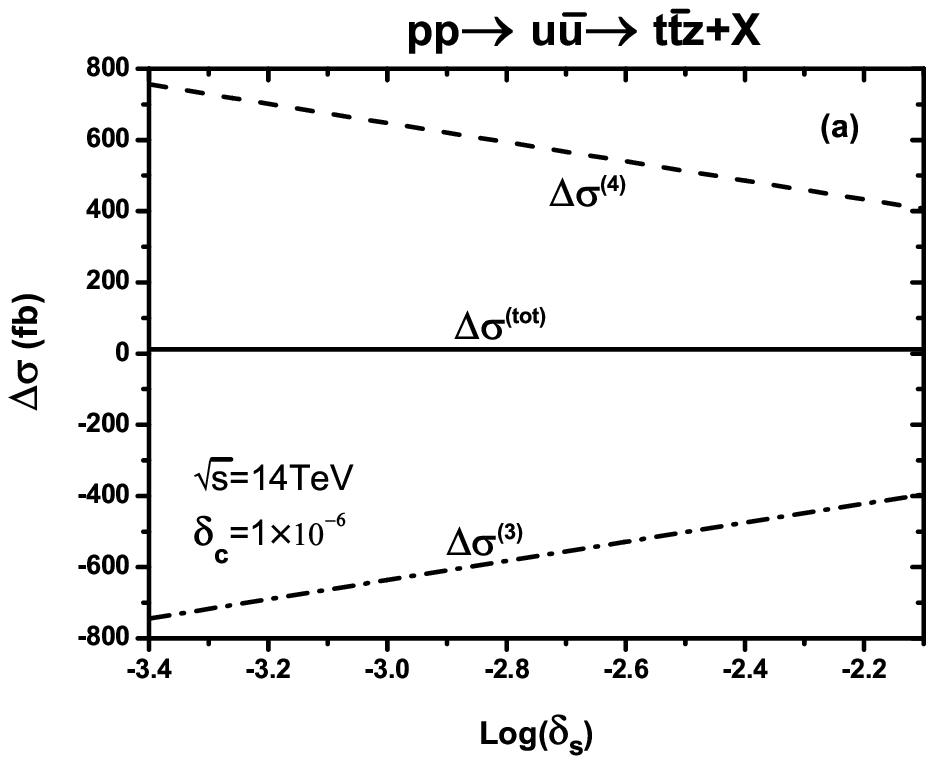}%
\hspace{0in}%
\includegraphics[width=3.2in,height=3in]{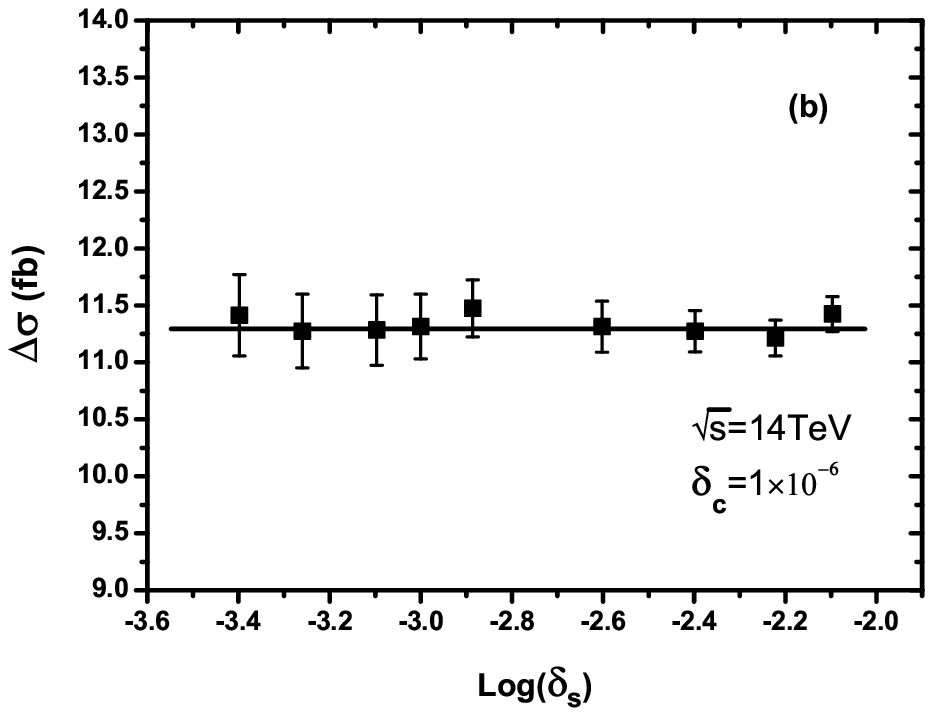}%
\hspace{0in}%
\caption{\label{fig7} (a) The dependence of QCD NLO correction
parts to the \ppuuttz process on the soft cutoff $\delta_s$ at the
LHC with the collinear cutoff $\delta_c =1 \times 10^{-6}$ and
$\sqrt{s}=14~TeV$. (b) The amplified curve for the total QCD
correction $\Delta\sigma_{SNLO}^{SM-like}$ to the process
\ppuuttz, where it includes the calculation errors. }
\end{figure}
\begin{figure}[htbp]
\includegraphics[width=3.2in,height=3in]{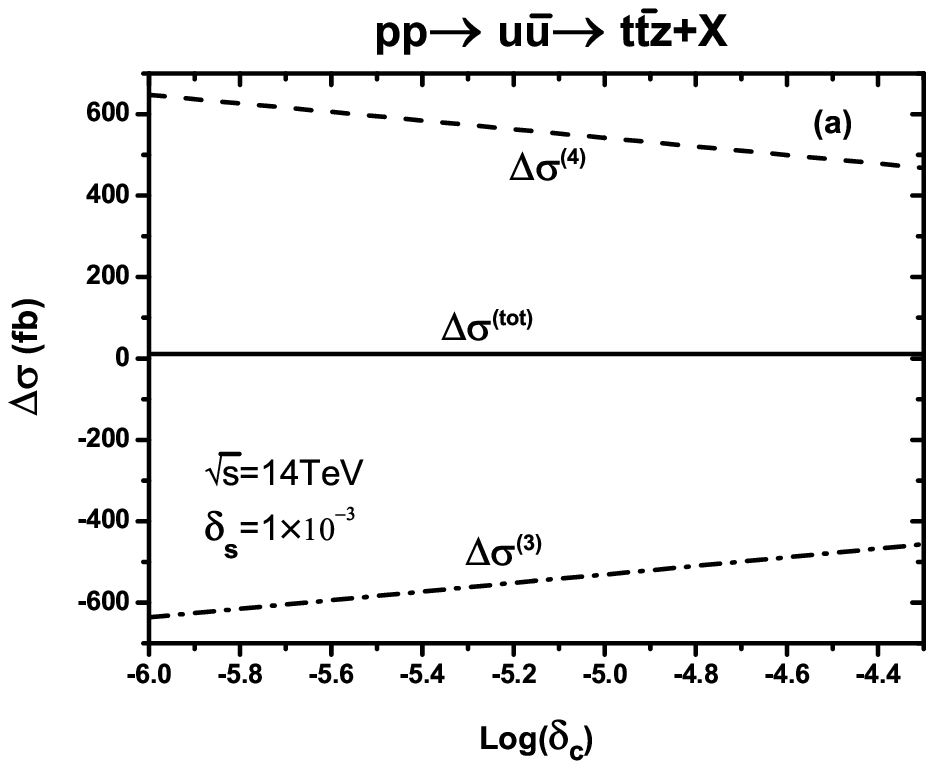}%
\hspace{0in}%
\includegraphics[width=3.2in,height=3in]{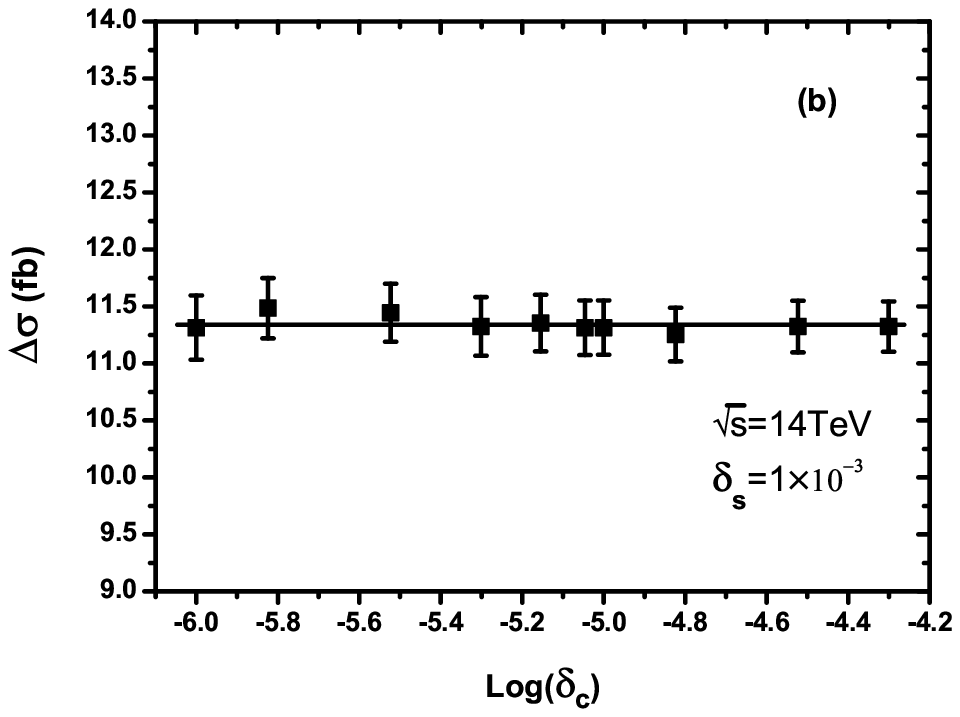}%
\hspace{0in}%
\caption{\label{fig8} (a) The dependence of the QCD NLO
correction parts to the \ppuuttz process on the collinear cutoff
$\delta_c$ at the LHC with $\delta_s=1\times 10^{-3}$ and
$\sqrt{s}=14~TeV$. (b) The amplified curve for the total QCD
correction $\Delta\sigma_{SNLO}^{SM-like}$ to the process
\ppuuttz, where it includes the calculation errors.}
\end{figure}

To make the demonstration of the correctness of our calculation for
the integrations over the phase space of the 4-body final-state real
emission processes, we have checked the independence of the SM-like
NLO QCD correction component $\Delta\sigma_{SNLO}^{SM-like}$ of the
process \ppuuttz on the two cutoffs $\delta_s$ and $\delta_c$
separately shown in Figs.\ref{fig7}(a-b) and
Figs.\ref{fig8}(a-b) separately. In identifying the
$\Delta\sigma_{SNLO}^{SM-like}$ independence on $\delta_s$, we fix
$\delta_c =1 \times 10^{-6}$ and vary $\delta_s$ from $4 \times
10^{-4}$ to $8 \times 10^{-3}$. For probing the
$\Delta\sigma_{SNLO}^{SM-like}$ independence on $\delta_c$, we take
$\delta_s =1 \times 10^{-3}$ and let $\delta_c$ running from $1
\times 10^{-6}$ to $4 \times 10^{-5}$. These four figures show that
although the three-body correction($\Delta\sigma^{(3)}$) and
four-body correction($\Delta\sigma^{(4)}$) are strongly related to
the cutoffs $\delta_s$ and $\delta_c$, the final total SM-like NLO
QCD correction $\Delta\sigma_{SNLO}^{SM-like}$ to the \ppuuttz
process, which is the summation of the three-body term and four-body
term, i.e.,
$\Delta\sigma_{SNLO}^{SM-like}=\Delta\sigma^{(3)}+\Delta\sigma^{(4)}$,
is indeed independent of the cutoffs ($\delta_s$ and $\delta_c$)
within the statistical errors.

\par
\begin{figure}
\centering
\includegraphics[scale=0.75]{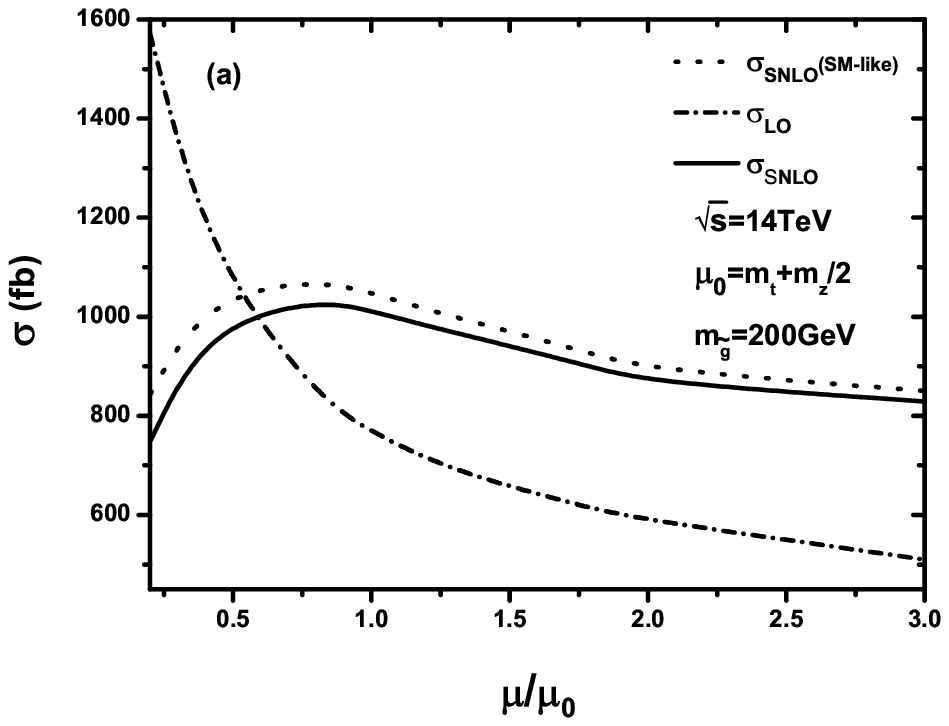}
\includegraphics[scale=0.75]{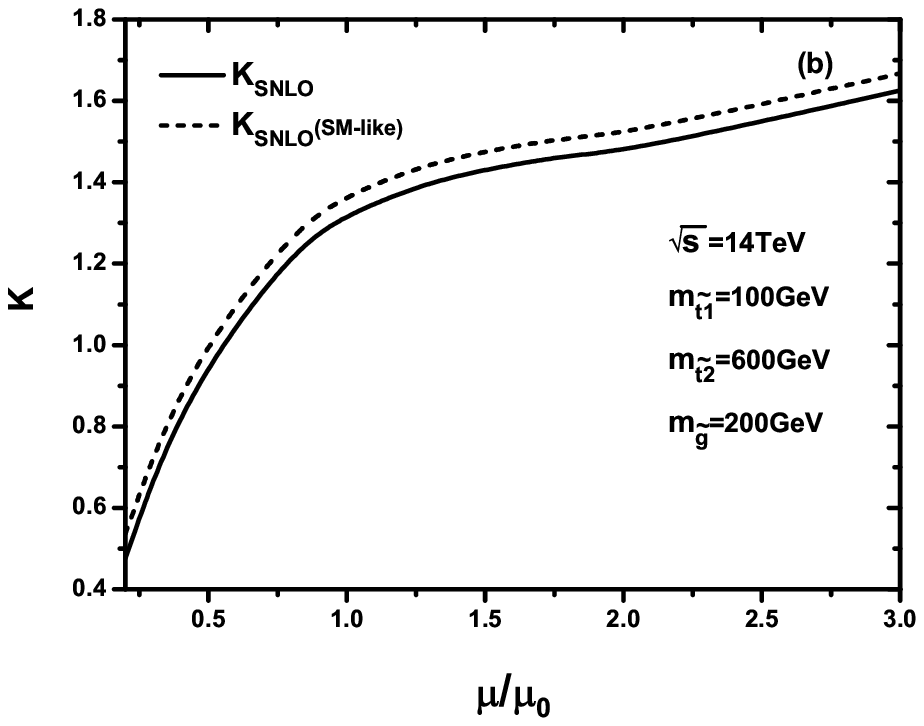}
\caption{\label{fig9a} (a) The dependence of the LO and NLO cross
sections in the CP-conserving MSSM and the SM on the energy scale at
the LHC. (b) The total NLO QCD K-factors for the
process($K_{SNLO}(\mu) \equiv \sigma_{SNLO}(\mu)/\sigma_{LO}(\mu)$)
and the NLO QCD K-factors of the SM-like part
($K_{SM-like}(\mu)=\sigma_{SNLO}^{SM-like}(\mu)/\sigma_{LO}(\mu)$)
for the \ppttz process versus the energy scale at the LHC.  }
\end{figure}

The LO and NLO QCD corrected cross sections in the CP-conserving
MSSM for the process \ppttz as the functions of the renormalization
and fatorization scales in the $\mu\equiv\mu_r=\mu_f$ way at the
LHC, are demonstrated in Fig.\ref{fig9a}(a). The curve for the
component of the SM-like NLO QCD corrected cross section is also
shown there for the comparison. The corresponding K-factors for the
NLO SQCD and SM-like QCD corrections, which are defined as $K(\mu)
\equiv \sigma_{SNLO}(\mu)/\sigma_{LO}(\mu)$ and
$K_{SM-like}(\mu)\equiv\sigma_{SNLO}^{SM-like}(\mu)/\sigma_{LO}(\mu)$,
are drawn in Fig.\ref{fig9a}(b), respectively. There we take $\mu_0
= m_t + m_Z/2$, $m_{\tilde{g}} = 200 GeV$ and $\mu$ running from
$\mu_0 / 5$ to $3 \mu_0$. As we know, the scale dependence of the
PDFs for the incoming $u$- and $d$-quarks are significant, even up
to the NLO the scale dependence of the results are not too small.
Since the $t\bar{t}Z^0$ production at the LHC is a QCD process at
the LO, the scale dependence of the NLO SM-like QCD corrected cross
section (dotted curve in Fig.\ref{fig9a}(a)) is less than that of
the LO cross section for the \ppttz process. On the other hand, the
pSQCD correction is to some extend a LO contribution, because this
process does not involve the supersymmetric strong coupling
$\hat{g}_s$ at the LO. Therefore, the pSQCD correction induces some
more scale dependence to the \ppttz process. Considering the fact
that pSQCD correction is quite small comparing with the SM-like QCD
correction demonstrated in Fig.\ref{fig9a}(a), the scale dependence
of the full SQCD corrected cross section is similar to that of the
SM-like QCD corrected cross section, which is much less than that of
the LO cross section. Actually, Fig.\ref{fig9a}(a) demonstrates that
the LO cross section is strongly correlated with the energy scale
$\mu$, while the NLO QCD corrections obviously improve the scale
uncertainties in both the CP-conserving MSSM and the SM. Comparing
with the SM-like NLO QCD correction, it can be seen the pSQCD
corrections cancellate the correction part from the SM-like QCD, and
the NLO pSQCD correction to the total cross section can exceed
$-4.75\%$ in our chosen parameters space. Fig.\ref{fig9a}(a) shows
that the total NLO SQCD K-factor changes from $0.48$ to $1.63$ as
the scale $\mu$ running from $\mu_0 / 5$ to $3 \mu_0$. In the
following calculations we set $\mu=\mu_0$.

\par
\begin{figure}
\centering
\includegraphics[scale=0.75]{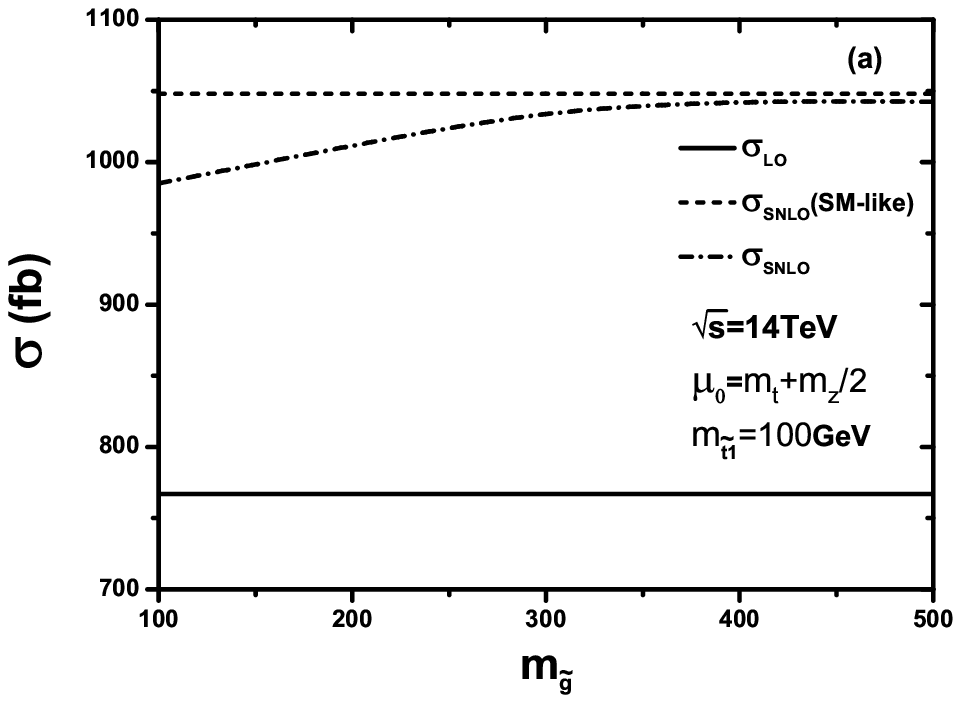}
\includegraphics[scale=0.75]{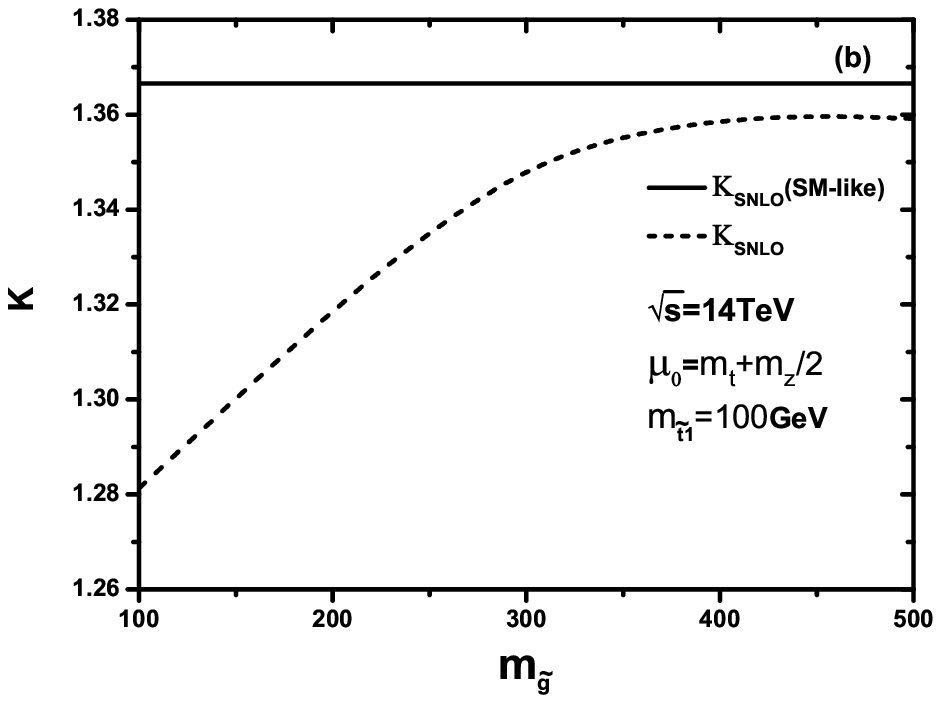}
\caption{\label{fig10a} (a) The LO and NLO QCD corrected cross
sections in the CP-conserving MSSM as the functions of the gluino
mass $m_{\tilde{g}}$ at the LHC. (b) The corresponding total NLO
QCD K-factor in the MSSM for the process \ppttz ($
K_{SNLO}(m_{\tilde{g}})\equiv
\frac{d\sigma_{SNLO}}{dm_{\tilde{g}}}/\frac{d\sigma_{LO}}{dm_{\tilde{g}}}$)
and the NLO QCD K-factor for the SM-like part for the \ppttz
process versus the gluino mass at the LHC.  }
\end{figure}

Fig.\ref{fig10a}(a) shows the LO and NLO QCD corrected cross sections
in the CP-conserving MSSM and the SM for the process \ppttz as the
functions of the gluino mass ($m_{\tilde{g}}$) at the LHC by taking
$m_{\tilde{t}_1} = 100 GeV$, $m_{\tilde{t}_2} = 600 GeV$, and the
mixing angle $\theta_{t}=-\pi/4$ for the stop sector. The
corresponding K-factors for the NLO SQCD and SM-like QCD corrections
versus $m_{\tilde{g}}$ are depicted in Fig.\ref{fig10a}(b),
respectively. Figs.\ref{fig10a}(a) and \ref{fig10a}(b) demonstrate
that although the SM-like curves have no relation with gluon mass,
the curves for NLO SQCD corrections is clearly related with
$m_{\tilde{g}}$ in the region of $m_{\tilde{g}}<300~GeV$. The NLO
SQCD corrected cross section (and K-factors) approaches a constant
when $m_{\tilde{g}}>400~GeV$ due to the decoupling effect for heavy
gluino exchanging. We can see that the NLO pSQCD correction is
non-zero when $m_{\tilde{g}}>400~GeV$, because of the relatively
light mass of $\tilde{t}_1$ in loops ($m_{\tilde{t}_1}=100~GeV$).
Fig.\ref{fig10a}(b) shows when we take $m_{\tilde{g}} = 100~GeV$, the
pSQCD relative correction can reach to $-8.56\%$ and K-factor of the
total SQCD correction will be $1.281$, while when we fix
$m_{\tilde{g}} = 200~GeV$, we get $-4.75\%$ for the corresponding
pSQCD relative correction and $1.319$ for the K-factor of the SQCD
correction.

\par
\begin{figure}
\centering
\includegraphics[scale=0.75]{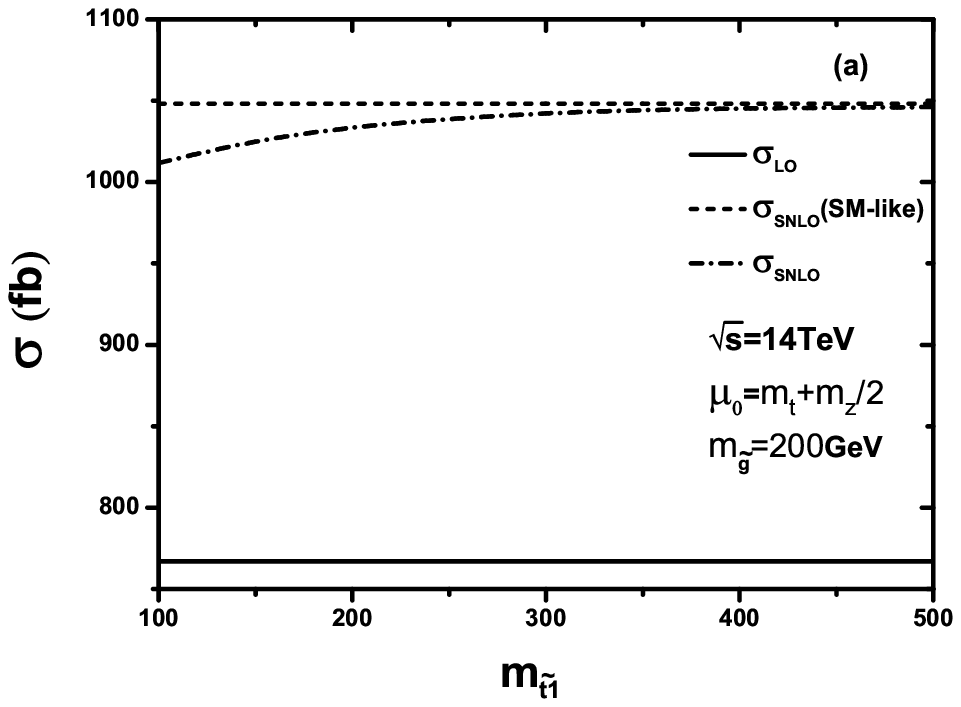}
\includegraphics[scale=0.75]{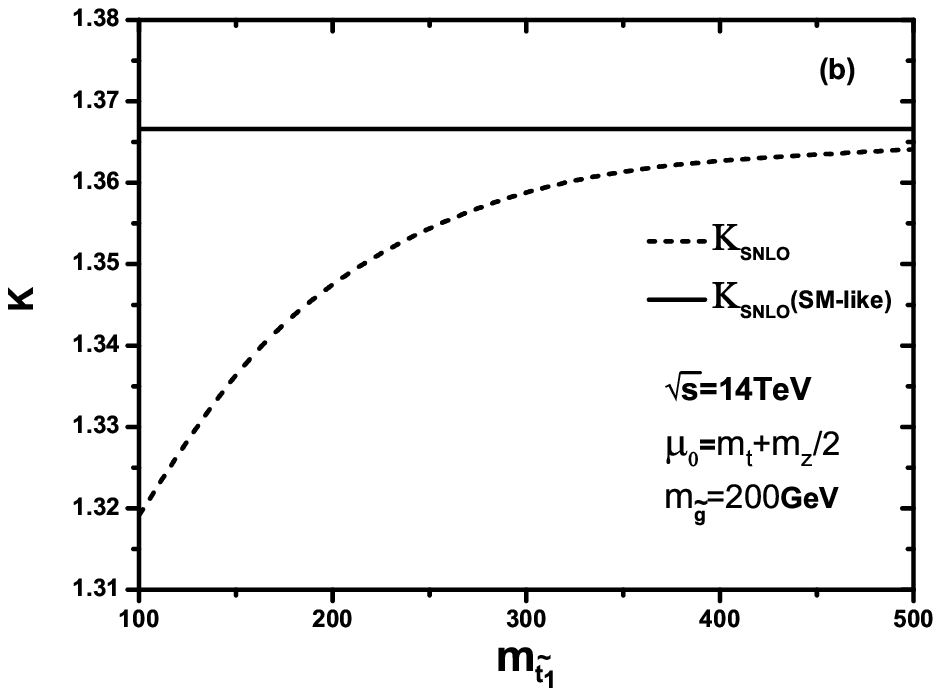}
\caption{\label{fig11a} (a) The LO and NLO QCD corrected cross
sections in the CP-conserving MSSM as the functions of the light
scalar top-quark mass $m_{\tilde{t}_1}$ at the LHC. (b) The
corresponding total NLO QCD K-factor in the MSSM  for the process
\ppttz ($K(m_{\tilde{t}_1})\equiv
\frac{d\sigma_{SNLO}}{dm_{\tilde{t}_1}}/\frac{d\sigma_{LO}}{dm_{\tilde{t}_1}}$)
and the NLO QCD K-factor for the SM-like part for the \ppttz
process versus $m_{\tilde{t}_1}$ at the LHC.  }
\end{figure}

We present Fig.\ref{fig11a}(a) to show the relations between the LO
and NLO QCD corrected cross sections in the CP-conserving MSSM and
SM for the process \ppttz at the LHC as the functions of
$m_{\tilde{t}_1}$, one of the input parameters for stop sector. Here
we take $m_{\tilde{g}}=200~GeV$, the other two input parameters for
stop sector are set as $m_{\tilde{t}_2} = 600 GeV$,
$\theta_{t}=-\pi/4$, and the other SUSY parameters are obtained as
explained above. The corresponding K-factors for the NLO SQCD and
SM-like NLO QCD corrections versus $m_{\tilde{t}_1}$ are depicted in
Fig.\ref{fig11a}(b), respectively. Again we see from
Figs.\ref{fig11a}(a) and (b) that the SM-like curves do not show the
dependence on $m_{\tilde{t}_1}$, but the NLO QCD corrections in the
MSSM obviously rely on $m_{\tilde{t}_1}$ in the region of
$m_{\tilde{t}_1}<400~GeV$. While in the region of
$m_{\tilde{t}_1}>400~GeV$ the NLO QCD corrections and corresponding
K-factors in the MSSM tend to be constant respectively, because
there exists the decoupling effect of heavy $\tilde{t}_1$ in loops,
and the nonzero NLO pSQCD correction is induced by the relatively
light mass of gluon in loops($m_{\tilde{g}}=200~GeV$). From
Fig.\ref{fig11a}(b) we can see that with the mass of $\tilde{t}_1$
running from $100~GeV$ to $500~GeV$, the NLO pSQCD relative
correction varies from $-4.75\%$ to $-0.25\%$, and the K-factor of
the total NLO SQCD changes from $1.319$ to $1.364$.

\par
\begin{figure}
\centering
\includegraphics[scale=0.7]{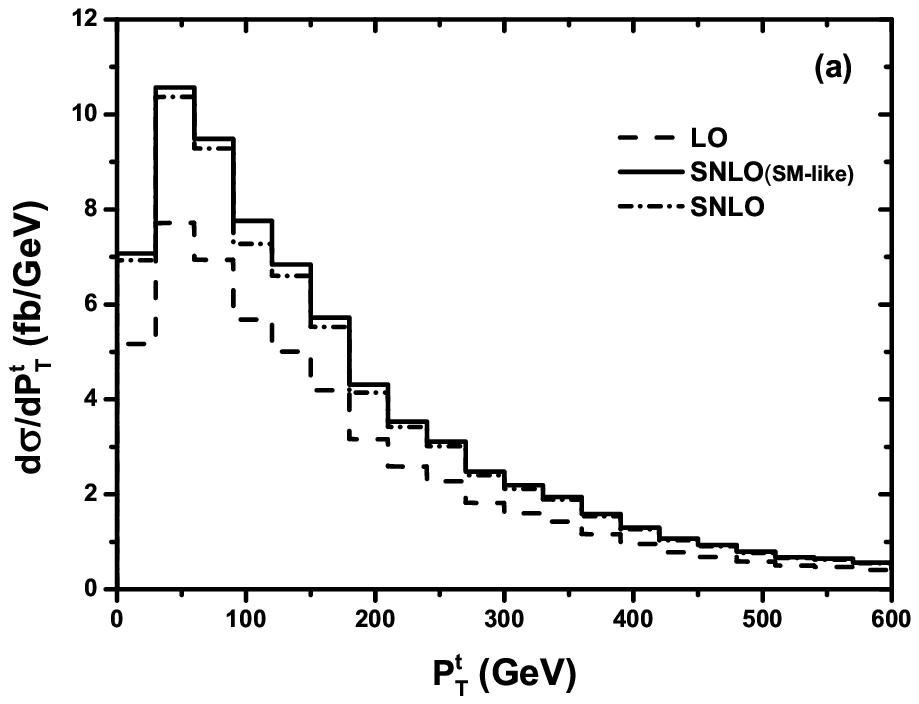}
\includegraphics[scale=0.7]{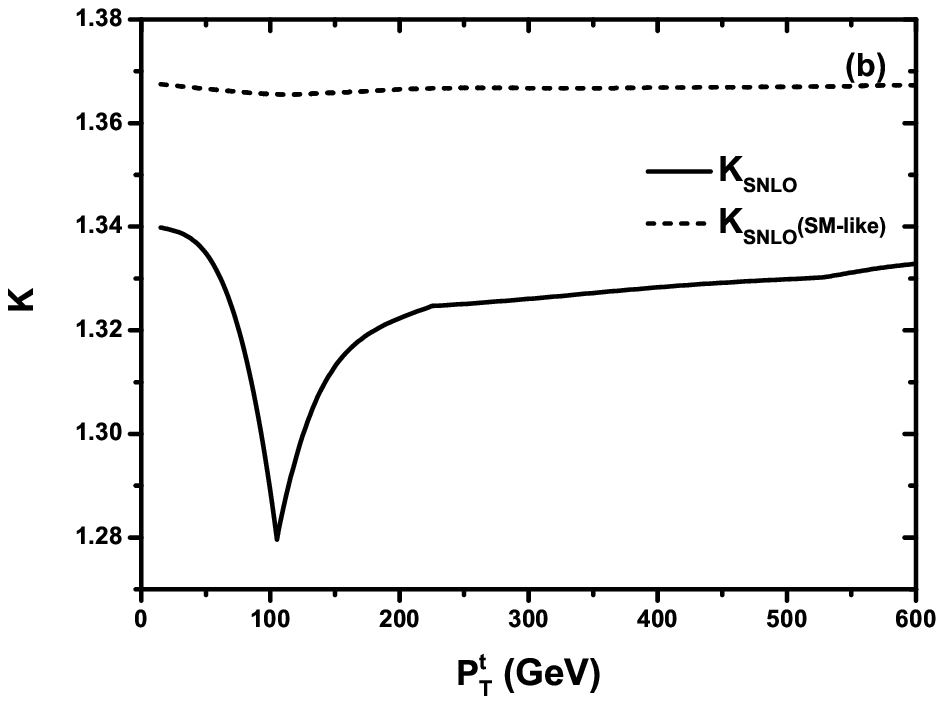}
\caption{\label{fig12} The LO and NLO QCD corrected distributions
of the transverse momentum of the top quark for the process \ppttz
in the SM and CP-conserving MSSM at the LHC and the corresponding
K-factors($K(p_T^{t})\equiv
\frac{d\sigma_{SNLO}}{dp_{T}^{t}}/\frac{d\sigma_{LO}}{dp_T^{t}}$)
versus $p_T^t$. (a) the differential cross section of the
transverse momentum, (b) the corresponding K-factors. }
\end{figure}
\begin{figure}
\centering
\includegraphics[scale=0.7]{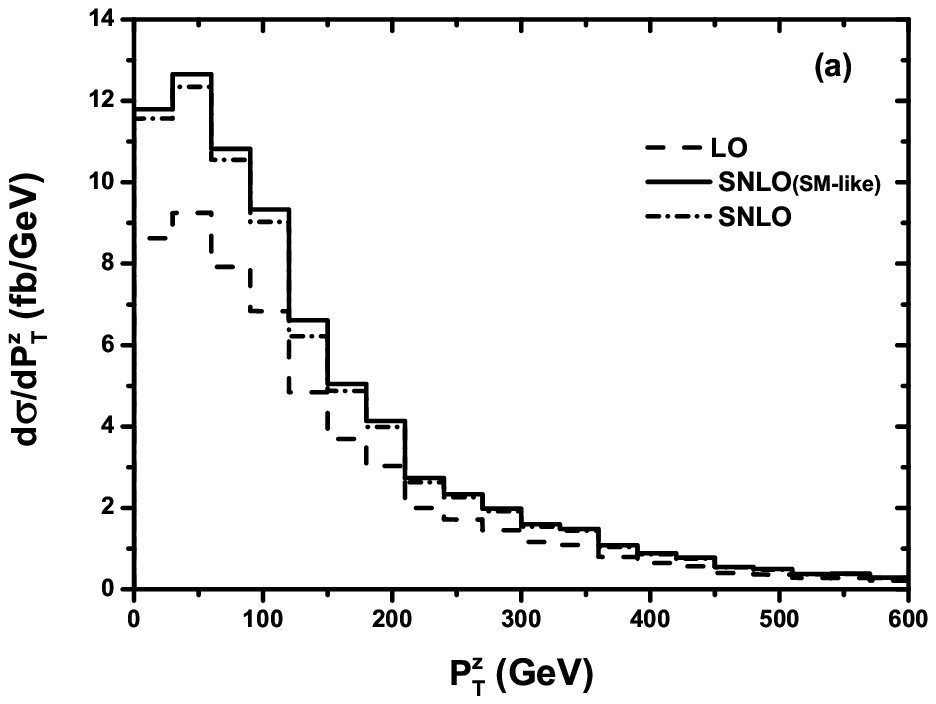}
\includegraphics[scale=0.7]{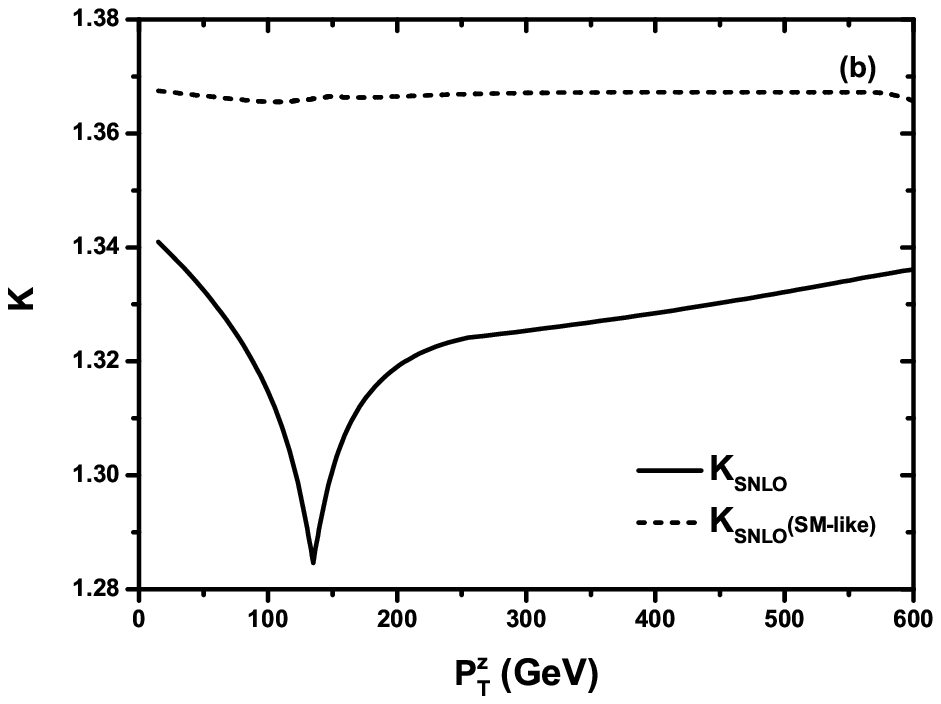}
\caption{\label{fig13} The LO and NLO QCD corrected distributions
of the transverse momentum of the $Z^0$-boson for the process
\ppttz in the SM and CP-conserving MSSM at the LHC and the
corresponding K-factors($K(p_T^{Z})\equiv
\frac{d\sigma_{SNLO}}{dp_{T}^{Z}}/\frac{d\sigma_{LO}}{dp_T^{Z}}$)
versus $p_T^Z$. (a) the differential cross section of the
transverse momentum, (b) the corresponding K-factors. }
\end{figure}

The LO and NLO QCD corrected differential cross sections of the
transverse momenta for top quark and $Z^0$ boson in the
CP-conserving MSSM and the SM for the process \ppttz at the LHC,
are drawn in Fig.\ref{fig12}(a) and Fig.\ref{fig13}(a),
separately. In these plots we take $m_{\tilde{t}_1} = 100 GeV$,
$m_{\tilde{t}_2} = 600 GeV$ and the mixing angle
$\theta_{t}=-\pi/4$ for stop sector, $m_{\tilde{g}}=200~GeV$, and
the other SUSY parameters are set to have the values explained
above. Their corresponding K-factors($K(p_T)\equiv
\frac{d\sigma_{SNLO}}{dp_T}/\frac{d\sigma_{LO}}{dp_T}$) are
depicted in Figs.\ref{fig12}(b) and Fig.\ref{fig13}(b),
respectively. There we take $m_{\tilde{g}}=200~GeV$. Both figures
Fig.\ref{fig12}(a) and Fig.\ref{fig13}(a) show that the SM-like
NLO QCD corrections enhance the differential cross sections of the
transverse momenta for the top quark and $Z^0$ boson in whole
plotted range, while the NLO pSQCD correction part suppresses the
SM-like QCD correction slightly. We can obtain from
Fig.\ref{fig12}(b) and Fig.\ref{fig13}(b) that the relative
corrections from the pSQCD can be $-8.56\%$ and $-8.12\%$ when
$p_T^{t} \sim 105~GeV$ and $p_T^{Z} \sim 135~GeV$ respectively.
And in these two figures there exist obvious distortions for the
two NLO SQCD K-factor curves compared with the corresponding
SM-like ones. Those curve distortions are caused by the resonant
effect of gluon self-energy.

\par
\begin{figure}
\centering
\includegraphics[scale=0.7]{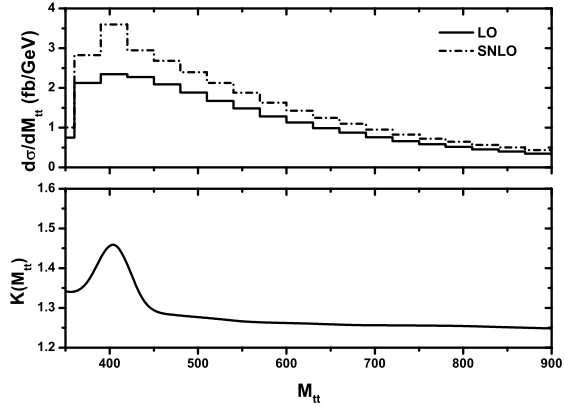}
\caption{\label{fig14} The LO and NLO QCD corrected differential
cross sections of the top-pair invariant mass $M_{(t\bar t)}$ and
the corresponding K-factors in the CP-conserving MSSM at the LHC.
}
\end{figure}

In Fig.\ref{fig14}, we define $K(M_{(t\bar
t)})=\frac{d\sigma_{SNLO}}{dM_{(t\bar
t)}}/\frac{d\sigma_{LO}}{dM_{(t\bar t)}}$ and plot the curves for
the differential cross sections and corresponding K-factor as the
functions of the top-pair invariant mass $M_{(t\bar t)}$, where we
take $m_{\tilde{t}_1} = 100 GeV$, $m_{\tilde{t}_2} = 600 GeV$,
$\theta_{t}=-\pi/4$ for stop sector, and $m_{\tilde{g}}=200~GeV$.
There we see the $K(M_{(t\bar t)})$ distribution demonstrates the
characteristic effects, which are shown already on the K-factor
curves in Fig.\ref{fig12}(b) and Fig.\ref{fig13}(b). The distortion
of the K-factor distribution curve for $M_{(t\bar t)}$ is located at
the vicinity of $M_{(t\bar t)}\sim 2 m_{\tilde{g}}=400~GeV$ where
the K-factor of the NLO SQCD correction reaches the value of
$1.457$. It exhibits exactly that the resonance effect of the
gluino-pair threshold in the gluon self-energy induces the curve
distortion.

\par
\begin{figure}
\centering
\includegraphics[scale=0.7]{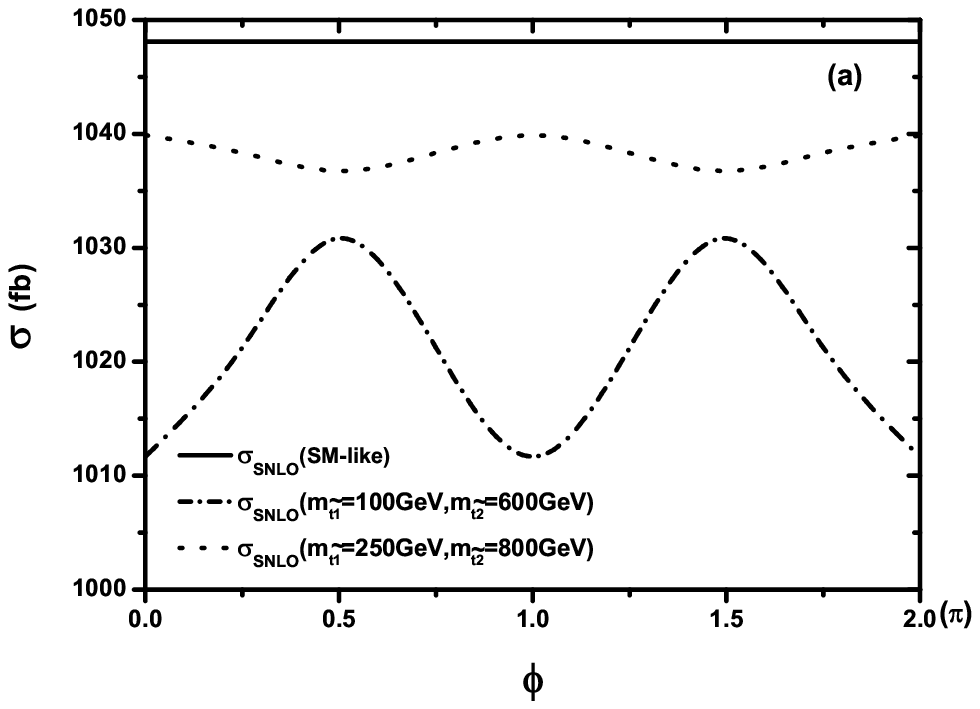}
\includegraphics[scale=0.7]{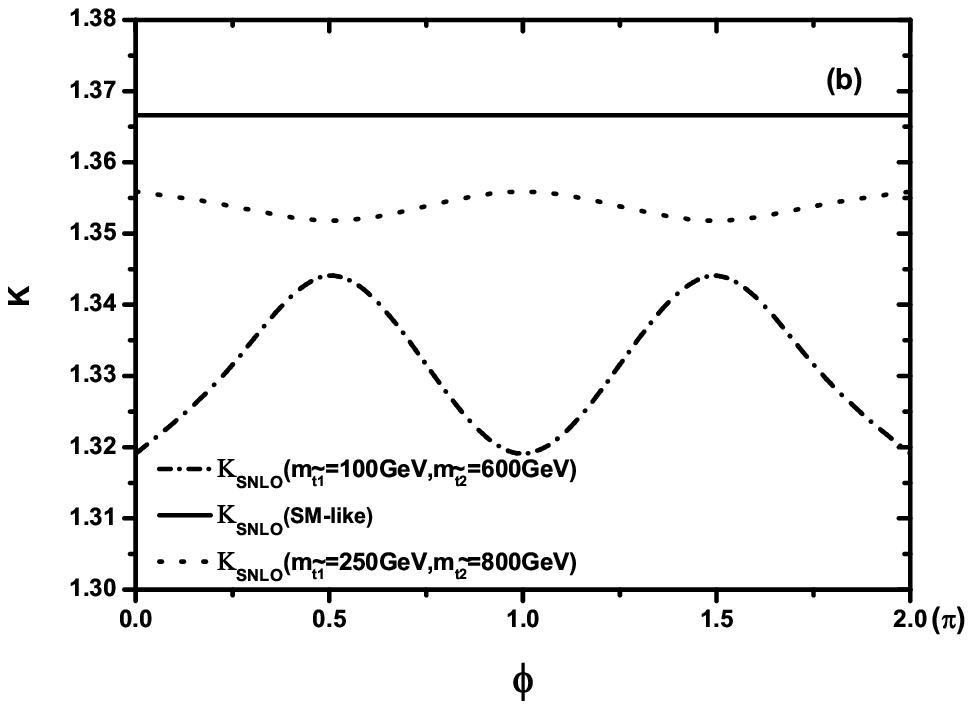}
\caption{\label{fig15}  (a) The LO and NLO QCD corrected cross
sections as the functions of the CP-phase $\phi(\equiv\phi_t)$ for
the process \ppttz at the LHC. (b) The corresponding total NLO QCD
K-factors versus CP phase $\phi$. }
\end{figure}

If the CP-violating MSSM is true, the CP-odd effects for the process
\ppttz at the LHC would be demonstrated through the cross section
deviation from the CP-conserving MSSM, and shown from a nonzero
CP-odd observable ${\cal A}_{\Phi}$ defined in
Eq.(\ref{CP-parameter}). We plot the NLO SQCD corrections to the
cross sections of \ppttz process as the functions of the CP-phase
$\phi$ in Fig.\ref{fig15}(a), and the corresponding K-factors are
drawn in Fig.\ref{fig15}(b). In these two figures we set
$\phi\equiv\phi_{t}$ and $\phi_q = 0,~(q=u,d,s,c,b)$,
$m_{\tilde{g}}=200~GeV$, for the stop sector we take two set of
input parameters: (1) $\{m_{\tilde{t}_1},m_{\tilde{t}_2},
\theta_{t}\}$
$=\{100GeV,600GeV,-\frac{\pi}{4}\}$(dash-dotted-curve); (2)
$\{m_{\tilde{t}_1},m_{\tilde{t}_2},\theta_{t}\}$
$=\{250GeV,800GeV,-\frac{\pi}{4}\}$(dotted-curve), and the other
SUSY parameter values are explained above. The curve for the SM-like
NLO QCD correction(full-line) in Fig.\ref{fig15}(a) shows the SM-like
correction part does not depend on the CP-phase, while the curves
for NLO SQCD correction part vary as cosine wave of $\phi$. The
curves for K-factors of the SM-like QCD and the SQCD correction in
Fig.\ref{fig15}(b), show the similar behaviors with the corresponding
ones in Fig.\ref{fig15}(a).

\par
\begin{figure}
\centering
\includegraphics[scale=0.8]{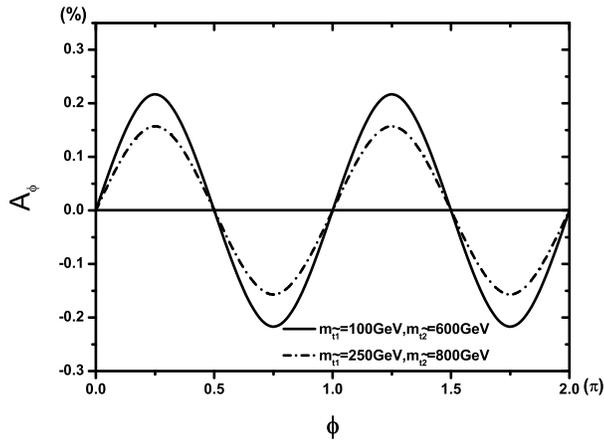}
\caption{\label{fig16}  The CP-violating asymmetry parameter
${\cal A}_{\Phi}$ as the functions of the CP-phase angle $\phi$
for the process \ppttz at the LHC, where the solid curve is for
$\{m_{\tilde{t}_1},m_{\tilde{t}_2},
\theta_{t}\}=\{100GeV,600GeV,-\pi/4\}$, and the dashed curve for
$\{m_{\tilde{t}_1},m_{\tilde{t}_2},
\theta_{t}\}=\{250GeV,800GeV,-\pi/4\}$. }
\end{figure}

We adopt the CP-asymmetry parameter(${\cal A}_{\Phi}$) definition in
Eq.(\ref{CP-parameter}) and depict ${\cal A}_{\Phi}$ as the function
of CP-phase $\phi$ in Fig.\ref{fig16}, where $m_{\tilde{g}}=200~GeV$
and the solid curve is for taking
$\{m_{\tilde{t}_1},m_{\tilde{t}_2}, \theta_{t}\}$
$=\{100GeV,600GeV,-\frac{\pi}{4}\}$, the dash-dotted curve
$\{m_{\tilde{t}_1},m_{\tilde{t}_2},\theta_{t}\}$
$=\{250GeV,800GeV,-\frac{\pi}{4}\}$. It shows that the two curves
for ${\cal A}_{\Phi}$ are sinusoidal with respect to $\Phi$, and the
absolute value of ${\cal A}_{\Phi}$ for
$\{m_{\tilde{t}_1},m_{\tilde{t}_2},
\theta_{t}\}=\{100GeV,600GeV,-\frac{\pi}{4}\}$ can reach its maximum
of $2.17\times 10^{-3}$ when $\phi=\frac{n\pi}{4},(n=1,3,5,7)$,
while it can be $1.57\times 10^{-3}$ for
$\{m_{\tilde{t}_1},m_{\tilde{t}_2},\theta_{t}\}
=\{250GeV,800GeV,-\frac{\pi}{4}\}$. From Eq.(\ref{Luminosity}) we
can see that if we assume the total cross section of the process
\ppttz is $1~pb$ and the integral luminosities are larger than
$212~fb^{-1}$, $1911~fb^{-1}$ or $5309~fb^{-1}$ we may observe the
CP-violating effect induced by $\phi$ at $1\sigma$, $3\sigma$ or
$5\sigma$ significance for the case of
$\{m_{\tilde{t}_1},m_{\tilde{t}_2},
\theta_{t}\}=\{100GeV,600GeV,-\frac{\pi}{4}\}$, and $406~fb^{-1}$,
$3651~fb^{-1}$ or $10142~fb^{-1}$ for the case of
$\{m_{\tilde{t}_1},m_{\tilde{t}_2},\theta_{t}\}
=\{250GeV,~800GeV,-\frac{\pi}{4}\}$, respectively.

\par
\section{Conclusion }
\par
In this paper, we study the total NLO QCD corrections for the
process \ppttz in the MSSM at the LHC. Our results show that both
the total NLO QCD corrections in the CP-conserving MSSM and the SM
improve the LO scale uncertainty. We provide the NLO QCD corrected
distributions of transverse momenta of the top-quark and $Z^0$ boson
at the LHC in the CP-conserving MSSM and the SM. There we can see
that the total NLO QCD corrections can modify significantly the LO
cross sections respectively. The pSQCD corrections to the process
\ppttz can be beyond $-8\%$ when we take $m_{\tilde{g}}=200~GeV$ and
restrict the top-quark $90 GeV < p_T^{t} < 120~GeV$ or the $Z^0$
boson $120~GeV < p_T^{Z} < 150~GeV$. And we see that the K-factor is
sensitive to the value of $m_{\tilde{g}}$ or $m_{\tilde{t}_1}$ in
the relatively lighter mass region of $\tilde{g}$ or $\tilde{t}_1$.
Furthermore, we find that if the CP-violating phase really exists in
the scalar top mixing matrix or in the Majorana mass term of the
gluino predicted by the CP-violating MSSM, the CP-violating effect,
described by CP-violating asymmetry parameter ${\cal A}_{\Phi}$, can
be expected to be the order of $10^{-3}$ and reach the maximal value
$2.17 \times 10^{-3}$. Therefore, testing CP-violation induced by
CP-phase $\phi$ in \ppttz process could be an interesting task at
the LHC.

\vskip 5mm
\par
\noindent{\large\bf Acknowledgments:} This work was supported in
part by the National Natural Science Foundation of
China(No.10875112, No.10675110), the Specialized Research Fund for
the Doctoral Program of Higher Education(No.20093402110030), and the
China Postdoctoral Science Foundation(No.20080440103).

\end{document}